\documentclass{article}

\usepackage{arxiv}
\usepackage{graphicx,subfigure}
\usepackage{cite}
\usepackage{mdframed}
\usepackage{empheq,mathtools}
\usepackage{amsmath,systeme}
\usepackage{mathrsfs}
\graphicspath{ {./images/}}
\usepackage[utf8]{inputenc} 
\usepackage[T1]{fontenc}    
\usepackage{hyperref}       
\usepackage{url}            
\usepackage{booktabs}       
\usepackage{amsfonts}       
\usepackage{nicefrac}       
\usepackage{microtype}      
\usepackage{lipsum}
\usepackage[nodayofweek,level]{datetime}

\title{Epidemic analysis of COVID-19 in Brazil by a generalized SEIR model}

\author{Suzete  Maria Silva Afonso\\
      Universidade Estadual Paulista (UNESP)\\
      Instituto de Geoci\^{e}ncias e Ci\^{e}ncias Exatas, 13506-900, Rio Claro-SP, Brasil. \\
      \texttt{s.afonso@unesp.br} \\
  \AND
  Juarez dos Santos Azevedo\\
  Universidade Federal  da Bahia (UFBA) \\
  Instituto de Ci\^{e}ncias, Tecnologia e Inova\c c\~ao,  Centro, 42802-721, Cama\c cari-BA, Brazil \\
  \texttt{jdazevedo@ufba.br} \\
   \And
Mariana Pinheiro Gomes da Silva\\
Universidade Federal do Rec\^oncavo da Bahia (UFRB)\\
Centro de Ci\^{e}ncias Exatas e Tecnol\'ogicas, Centro, 44380-000, Cruz das Almas-BA, Brazil.\\
\texttt{mpinheiro@ufrb.edu.br}
}

\begin{document}
\maketitle
\begin{abstract}

We shall apply a generalized SEIR model to study the outbreak of COVID-19 in  Brazil. In particular, we would like to explain the projections of the increase in the level of infection over a long period of time, overlapping large local outbreaks in the most populous states in the country. A time-dependent dynamic SEIR model inspired in a model previously used during the outbreak in China was used to analyses the time trajectories of infected, recovered, and deaths. The model has parameters that vary with time and are fitted  considering  a nonlinear least-squares method. The simulations starting from April 8, 2020, concluded that  the time for a peak in  Brazil will be in  July 21, 2020 with total cumulative infected cases around 982K people; in addition, an estimated total death case will reach to 192K in the end. Besides that, Brazil will reach a peak in terms of daily new infected cases and death cases around the middle of July  with 50K cases of infected and almost 6.0K daily deaths.
\end{abstract}


\section{Introduction.}
In recent times, COVID-19 has become one of humanity's biggest epidemic problems, gaining pandemic status and, having a significant impact on mortality rates in both rich and poor countries. The current balance (May 18, 2020) of World Health Organization (WHO) records that a total of around 5 million people were infected by the COVID-19 causing 320K deaths \cite{WHO}.  In Brazil, since the identification of the first case of COVID-19, in February, 2020, approximately 310 thousand cases of the disease have already been identified and almost 20 thousand deaths (Ministry of Health of Brazil, in the last epidemiological bulletin at (May 18, 2020)) \cite{MinNT}. Currently, this epidemic is one of the main public health problems in the world. The alarming numbers of this epidemic have encouraged several authors to work on epidemic models that can predict the progress of this disease.

Some mathematical models are proposed for studying the spread of viruses among humans, such as SIR (Susceptible, Infected, Recovered) model \cite{bjornstad2002dynamics, ji2014threshold, toda2020susceptible, zhang2013stochastic}
and  SEIR (Susceptible, Exposed, Infected, Recovered) model \cite{SEIR, Chenseirmodel,Weiss, modeloEIR}. The SEIR model derives from the SIR model and  require a large amount of data over time to execute.
Here, we follow the adapted SEIR model developed  by Tan and Chen in  \cite{tan2020realtime}, which is the most reliable response to emergency public health actions and takes into account the population exposed to the virus.

More specifically, we use the generalized version of the classical SEIR known as SEIRDP (Susceptible, Exposed, Infected, Recovered, Death, Insusceptible (P)) which is a particular case of the model proposed by \cite{peng2020epidemic}. In this model it is possible to include key epidemic parameters for COVID-19, such as the latent time, infected time, protection rate and time evolution of the recovery. This allows us to estimate the inflection point, ending time and total infected cases in Brazil and its states according to the Ministry of Health of Brazil data. The data are estimated considering a time window for collecting cure and mortality rates, which provides more accurate information on the evolution of local cases.


The search for a reliable estimate of the number of infected and deaths is crucial so that public health systems can act to respond quickly to the needs presented. For this reason, many studies aim to estimate the contagion curve, the spreading speed and the prediction of deaths.
Recently, Imperial College \cite{Imperial} released a study on the current situation of COVID-19 in Brazil and estimate that human-to-human transmission of Sars-CoV-2, virus of the coronavirus family causing the disease COVID-19, in Brazil averages from 1 to 2.8 people. They use a semi-deterministic Bayesian hierarchical model that analyzes the impacts of interventions aimed at restricting the transmission of the virus but it is not clear how effective these measures were. The model adopted follows a function of patterns in urban mobility to estimate deaths, infections, and transmissions. Some models can also predict the evolution of contaminants in real-time, such as the IHME \cite{IHME} with some important variables such as the transmission rate and the recovery rate, neither static nor variable over time, which makes the result very broad and somewhat vague.

Local health policies and actions can also be taken to mitigate the virus. Some authors have also studied the impact of lockdown on the spread of the virus \cite{lockdownhubei}. Others, who used these models, consider the study of public policies to mitigate the disease in specific locations \cite{lockdownhubeiwuhan}. Our simulations are based on general data from Brazil and where different policies have been applied in each state according to hospital capacity and speed of contamination. In this way, we focus on responses to general data from Brazil and in some Brazilian states with a higher rate of cases in order to assist in medical and political actions.

In this work, we analyzed the case data of COVID-19 from the states of São Paulo, Rio de Janeiro, Ceará, Espírito Santo, Bahia, Amazonas, Pará, Pernambuco, and Maranhão. To validate our results through the SEIRDP model, we carefully collect the epidemic data from the Ministry of Health of Brazil \cite{MinNT} from April 08th till  May 21th, 2020 and studied the approximation of the observed and computed data considering the infected, recovered and death cases.  Then,  we were able to estimate the parameters that characterize the natural history of this disease and allows us to investigate the projection of the epidemic in certain periods.

\section{Model and Methods}

\subsection{SEIRDP model}


The SEIRDP (Susceptible, Exposed, Infected, Recovered, Death, Insusceptible (P))  model used to analyze the coronavirus epidemic in Brazil has the following variables:
\begin{itemize}
	\item $S(t)$: Susceptible population;
	\item $E(t)$: Population who are exposed to the virus, but not yet infectious in latent period;
	\item $I(t)$: Population who get laboratory positive confirmation and with infectious capacity;
	\item $R(t)$: Recovery cases;
	\item $D(t)$: Death number;
	\item $P(t)$: Insusceptible cases;
	\item $N = S + P+ E + I + R +  D$ is the total population;
	\item $\alpha$: Protection rate (include people exposed to the infectious patients and people exposed to the asymptomatic patients);
	\item $\beta$:  Infection rate (include exposed people catch COVID-19,  people get COVID-19 with diagnosed confirmation and people; without symptoms but can transmit it to others);
	\item $\gamma^{-1}$: Average latent time;
	\item $\lambda(t)$: Coefficient used in the time-dependent cure rate;
	\item $\kappa(t)$: Coefficient used in the time-dependent mortality rate.
\end{itemize}

The variable described above are related through  the following ODE system:
\begin{equation}\label{md:SEIR}
\begin{split}
\frac{dS(t)}{dt}&=-\beta \frac{S(t)I(t)}{N}-\alpha S(t),\\
\frac{dE(t)}{dt}&=\beta \frac{S(t)I(t)}{N}-\gamma E(t), \\
\frac{dI(t)}{dt}&=\gamma E(t)-\lambda(t) I(t)-\kappa(t) I(t),\\
\frac{dR(t)}{dt}&=\lambda(t) I(t),\\
\frac{dD(t)}{dt}&=\kappa(t) I(t), \\
\frac{dP(t)}{dt}&=\alpha S(t).
\end{split}
\end{equation}

The model is time-dependent, because the cure rate $\lambda(t)$ and mortality rate $\kappa(t)$   are  analytical  functions of the time defined by
\begin{equation}\label{lambda}
\lambda(t) = \lambda_0/[1+ \exp(-\lambda_1 (t-\lambda_2))],
\end{equation}
\begin{equation}\label{kappa}
\kappa(t) = \kappa_0  \exp(-\kappa_1t),
\end{equation}
respectively, which are fitted  according to the number of cases for $t\in [t_0, t_f]$, where $t_0$ and $t_f$ are the initial and final periods, respectively.
Equations \eqref{lambda} and \eqref{kappa} indicate that cure rate or recovery rate
$\lambda(t)$ goes up exponentially toward a threshold value and the mortality
rate $\kappa(t)$  should be zero after an infinite time.
Moreover, the parameters $\{\alpha, \beta, \gamma^{-1}, \lambda(t), \kappa(t)\}$ were fitted  in the least square sense.
%
%
%

The parameters  $\{ \alpha$, $ \beta$, $\gamma^{-1}$, $\lambda_0$, $\lambda_1$,$\lambda_2$,
$\kappa_0(t)$,$\kappa_1(t)\}$ are computed simultaneously by a nonlinear least-squares solver~\cite{cheynet:2020}. We can replace them in the model \eqref{md:SEIR}  and calculate the  time-histories of the different states $\{S(t), P(t), E(t), I(t), \linebreak R(t), D(t)\}$.  In this case, we use the standard fourth--order Runge--Kutta process \cite{burden}. For this, the system of equations is written as a single ODE
\begin{equation}
\frac{d \mathbf{Y}}{d t} = \mathbf{A}\cdot\mathbf{Y}+\mathbf{F}
\end{equation}
such that
\begin{equation*}
\begin{split}
&\mathbf{Y}=
\begin{bmatrix}
S(t) & E(t) & I(t) & R(t) & D(t) & P(t)\\
\end{bmatrix}^T,\\
&\mathbf{A}=
\begin{bmatrix}
-\alpha & 0 & 0 & 0 & 0&0\\
0 & -\gamma & 0 & 0 & 0&0\\
0 & \gamma & -\lambda(t)-\kappa(t) & 0 & 0&0\\
0 & 0 & \lambda(t) & 0 & 0&0\\
0 & 0 & \kappa(t) & 0 & 0&0\\
\alpha & 0 & 0 & 0 & 0&0\\
\end{bmatrix},\\
&\mathbf{F}= S(t)\cdot I(t) \cdot
\begin{bmatrix}
-\frac{\beta}{N} & \frac{\beta}{N} & 0 & 0 & 0 & 0\\
\end{bmatrix}^T.\\
\end{split}
\end{equation*}

%
%
\section{Results}

Anchored by the information from the public data of the Ministry of Health of Brazil \cite{MinNT}, we reliably estimate key epidemic parameters and make projections on the inflection point considering fitted and actual data. Experiments show the spatial distribution of COVID-19 cases in  Brazil and in some states of the federation, namely, São Paulo, Ceará, and Rio de Janeiro. These states stand out for the highest incidence of  COVID-19 in Brazil. We also projected cases of COVID-19 in other states in the country: Espírito Santo, Bahia, Amazonas, Pará, Pernambuco and Maranhão that had a large number of confirmed cases in relation to other states in the federation, about $I>8.000$ each one, according in the last epidemiological bulletin of Ministry of Health of Brazil in May 18, 2020.

\subsection{Prediction of Time evolution of COVID-19 for Brazil}

\begin{figure}[htb]
	\centering 
	\subfigure[]{\label{fig_BrasilTotal}\includegraphics[scale=.5]{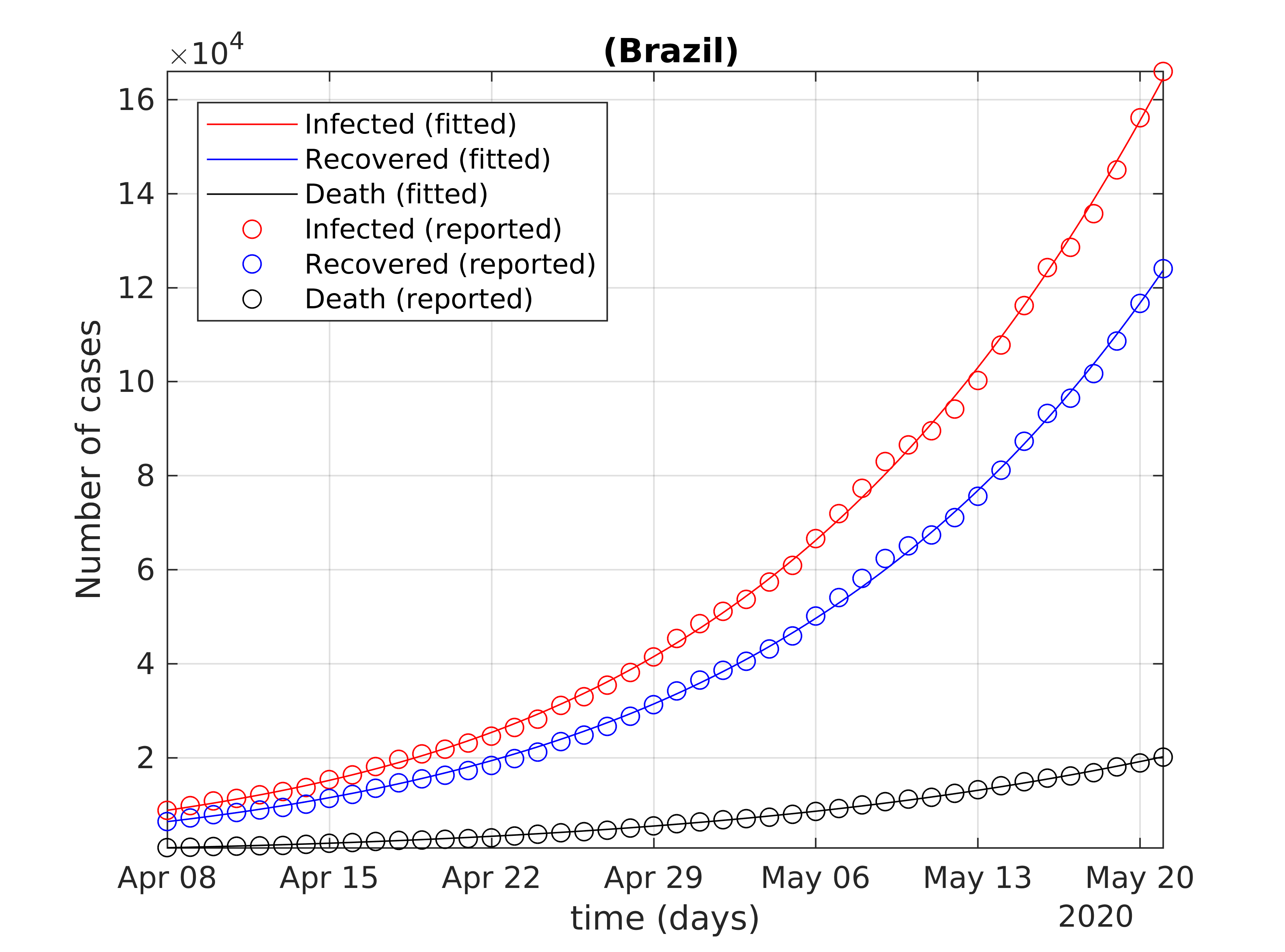}}
	\subfigure[]{\label{ErroRelative_Brasil}\includegraphics[scale=.5]{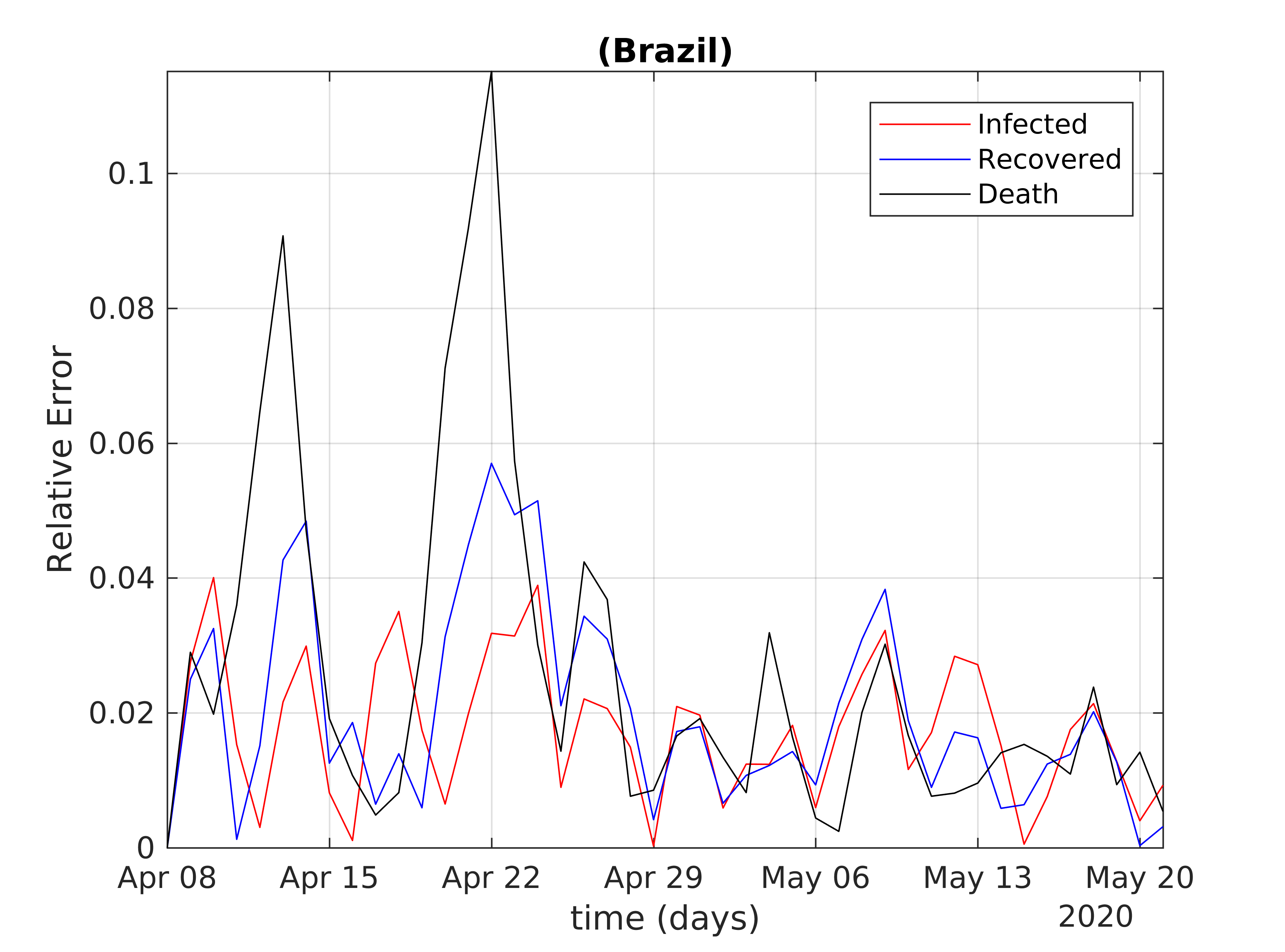}}\\
	\subfigure[]{\label{fig_BrasilTotal_SEIRVSSEIRDP}\includegraphics[scale=.5]{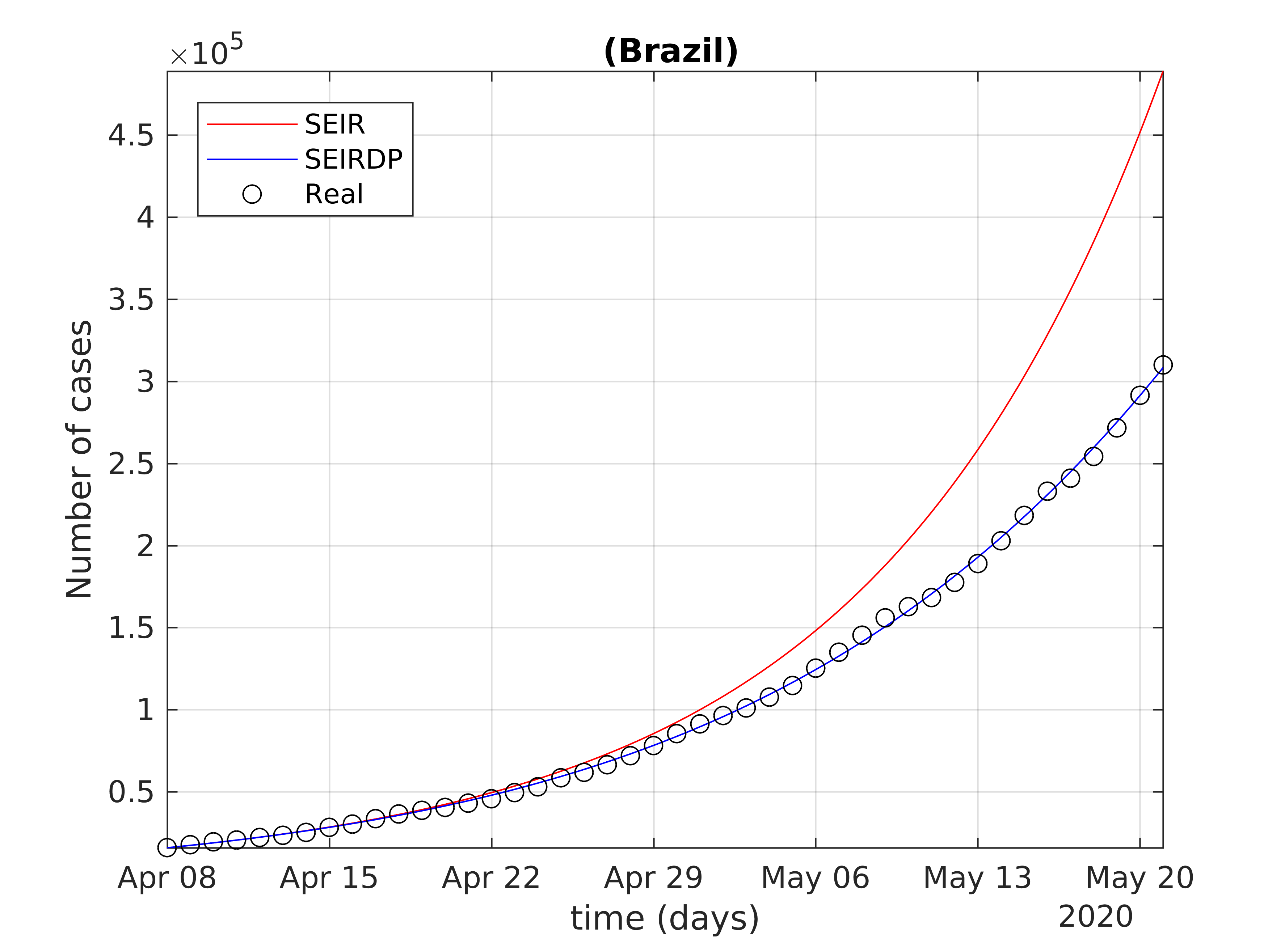}}	
	\caption{(a) Modeling of the differential SEIRDP model from April 8 to May 21. (b) The relative mean errors of projection for Brazil  between actual data and SEIRDP model involving cases of infected, recovered and death. (c) Modeling of the differential SEIR model from April 8 to May 21.}
	\label{fig_prev_BR}
\end{figure}

Fig.~\ref{fig_BrasilTotal} shows the first modeling and the projection for Brazil from late early April to mid-May 2020 for the currently total,
infected, recovery and death cases, while Fig. \ref{ErroRelative_Brasil} presents its respective relative error  solutions. According to these results we can say that the model has a good correspondence with the data provided. This means that the state parameters found by the least-squares solution should be suitable for projecting future dates. A test was done with the traditional SEIR and the results were overestimated (see Fig. \ref{fig_BrasilTotal_SEIRVSSEIRDP} ).

\begin{figure}[!ht]
	\centering
	\subfigure[]{\label{fig_BrasilPrev}\includegraphics[scale=.5]{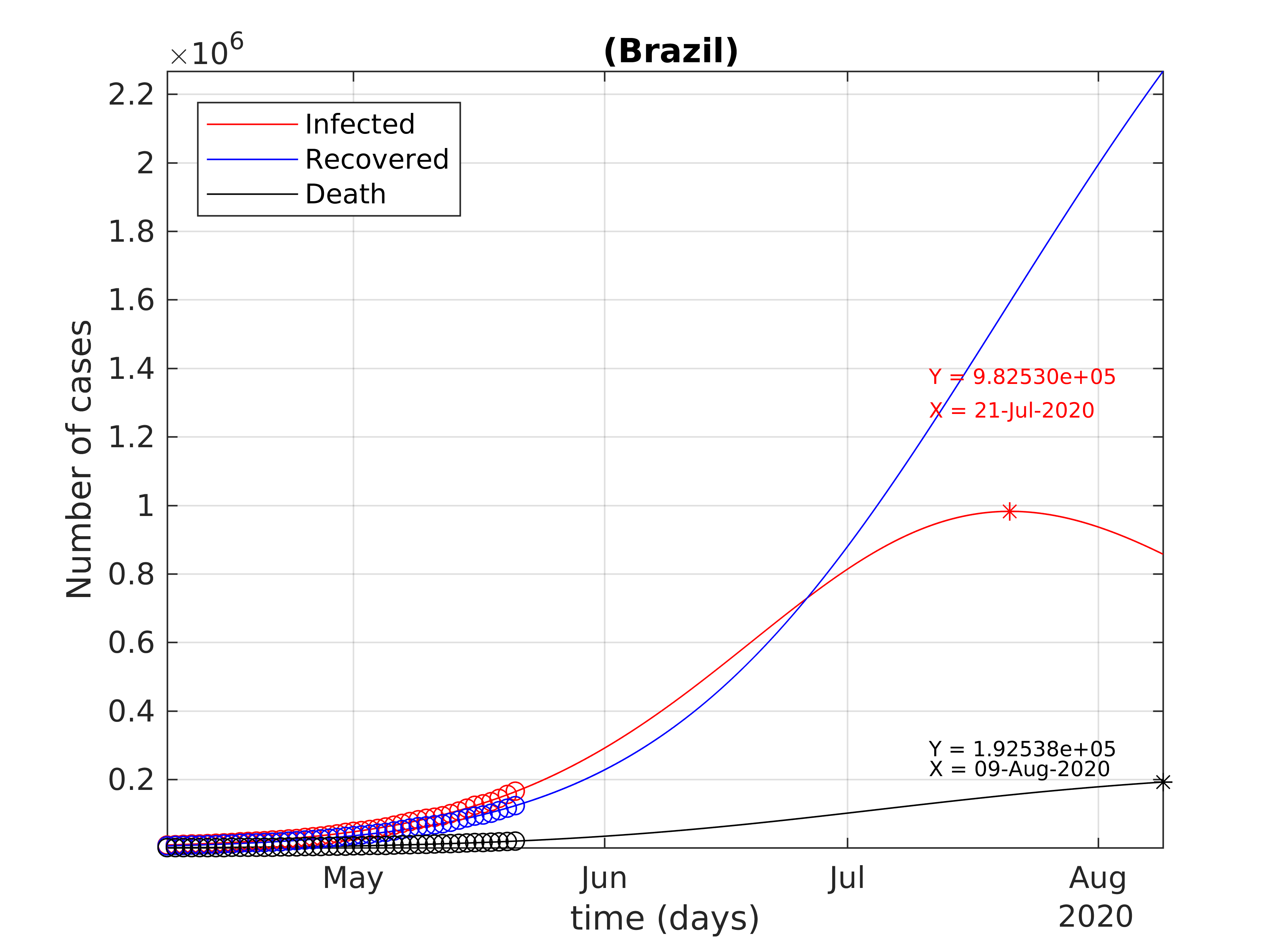}}\\
	\subfigure[]{\label{fig_BrasilConfirmed_D}\includegraphics[scale=.5]{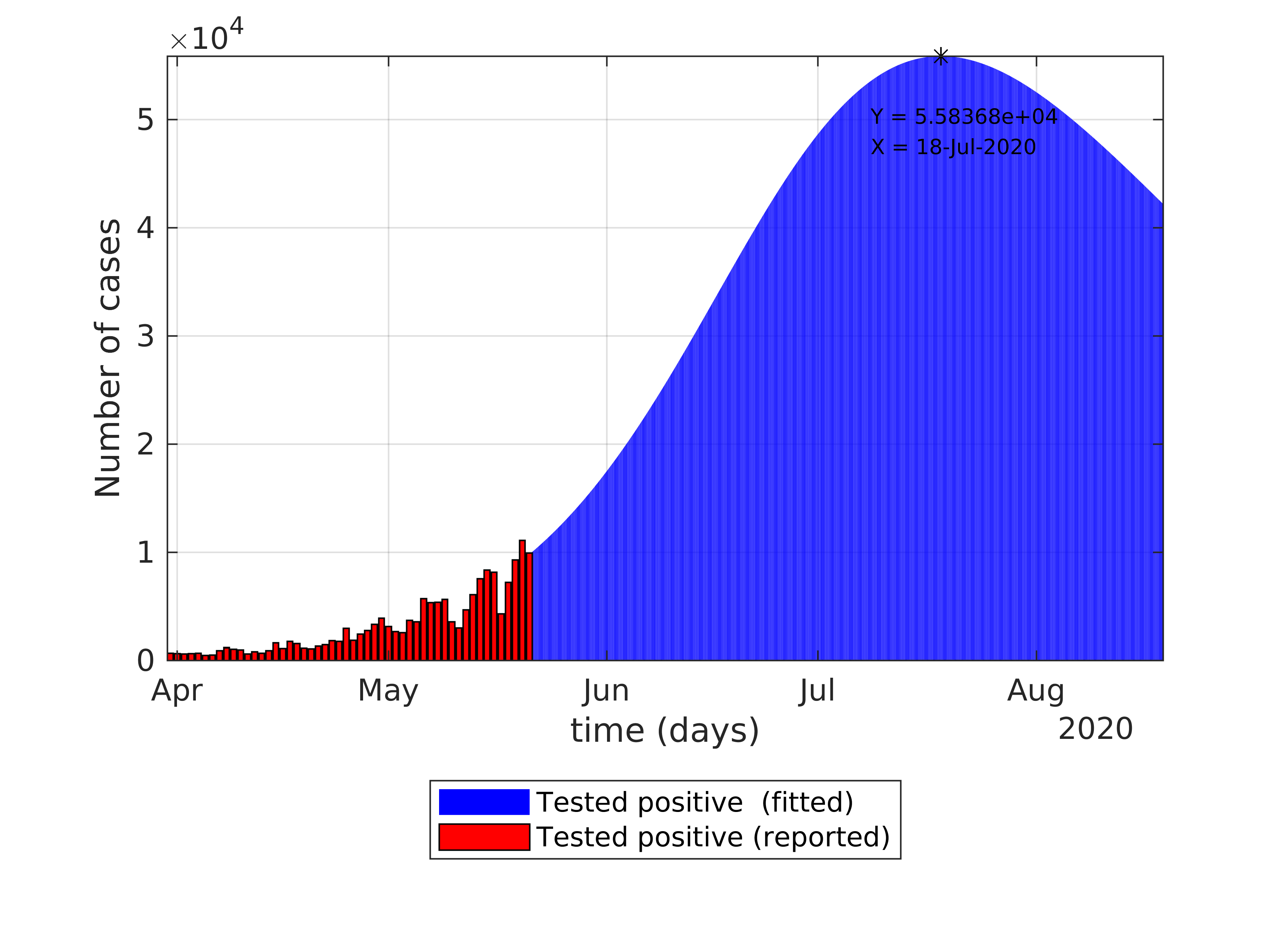}}
	\subfigure[]{\label{fig_BrasilDeath_2_D}\includegraphics[scale=0.5]{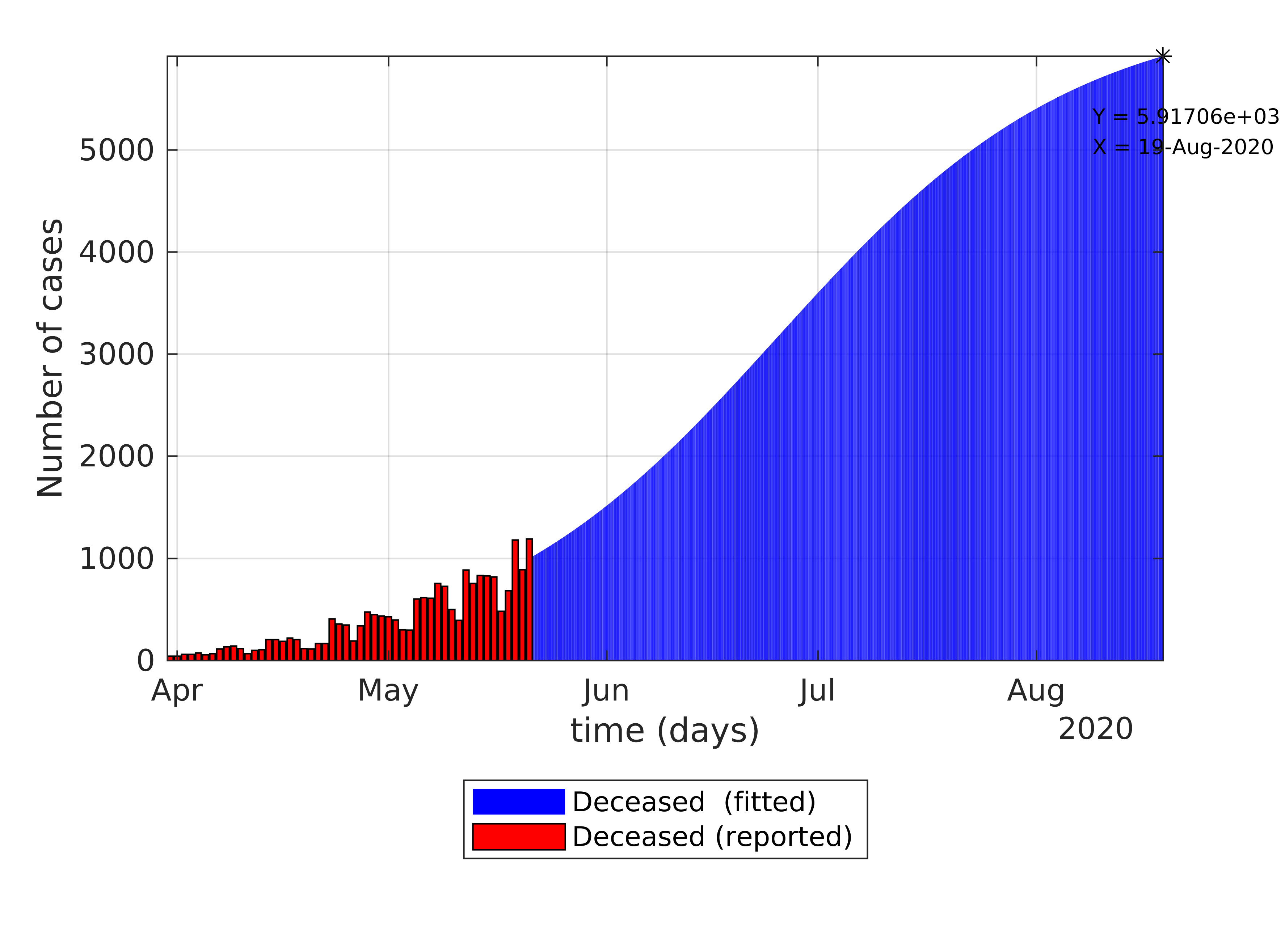}}
	\caption{(a) Predictions of the differential SEIRDP model for Brazil from	early April to final August, 2020. (b) The daily  of infected cases and (c) deceased measured and projected for Brazil early April to mid-August, 2020.}
	\label{fig:SEIRDP_BR_prev}
\end{figure}

In Fig. \ref{fig_BrasilPrev}, through the SEIRDP model, we made the projections in the period from April to August. According to this model, the infected cases in Brazil will reach to the peak  at middle of July 2020 with about 982K cases.  In this sense, the period also represents the peak time for medical resources get ready. According to the Ministry of Health, Brazil has a total of 441,811 hospital beds distributed in the country's numerous regions. From the number of infected, we can see that the public health system in Brazil may collapse in the middle of June.
The number of victims is also expected to grow in the same period. It should be noted that these results are only predictions based on data and population mitigation policies. These results may change as the population's isolation patterns change, as studies in this direction indicate, \cite{lockdownhubei, lockdownhubeiwuhan}. 

Fig.~\ref{fig_BrasilConfirmed_D} and Fig.~\ref{fig_BrasilDeath_2_D}
presents daily newly confirmed infected cases and daily death cases
for Brazil over the same period respectively. The red lines indicate the
actual data and the blue lines are predicted data from May to
early of August 2020.  As we can see that the daily infected cases
seem reaching the peak around July 21 with about 55K  per day and
then start to decrease.  From Fig.~\ref{fig_BrasilDeath_2_D}, we
can see that the daily death  is converging to approximately 6000 daily deaths  without a projection of decrease in this period.

\subsection{Prediction of Time evolution of COVID-19  for S\~ao Paulo state}

\begin{figure}[htb]
	\subfigure[]{\label{fig_StateBr1}\includegraphics[scale=.5]{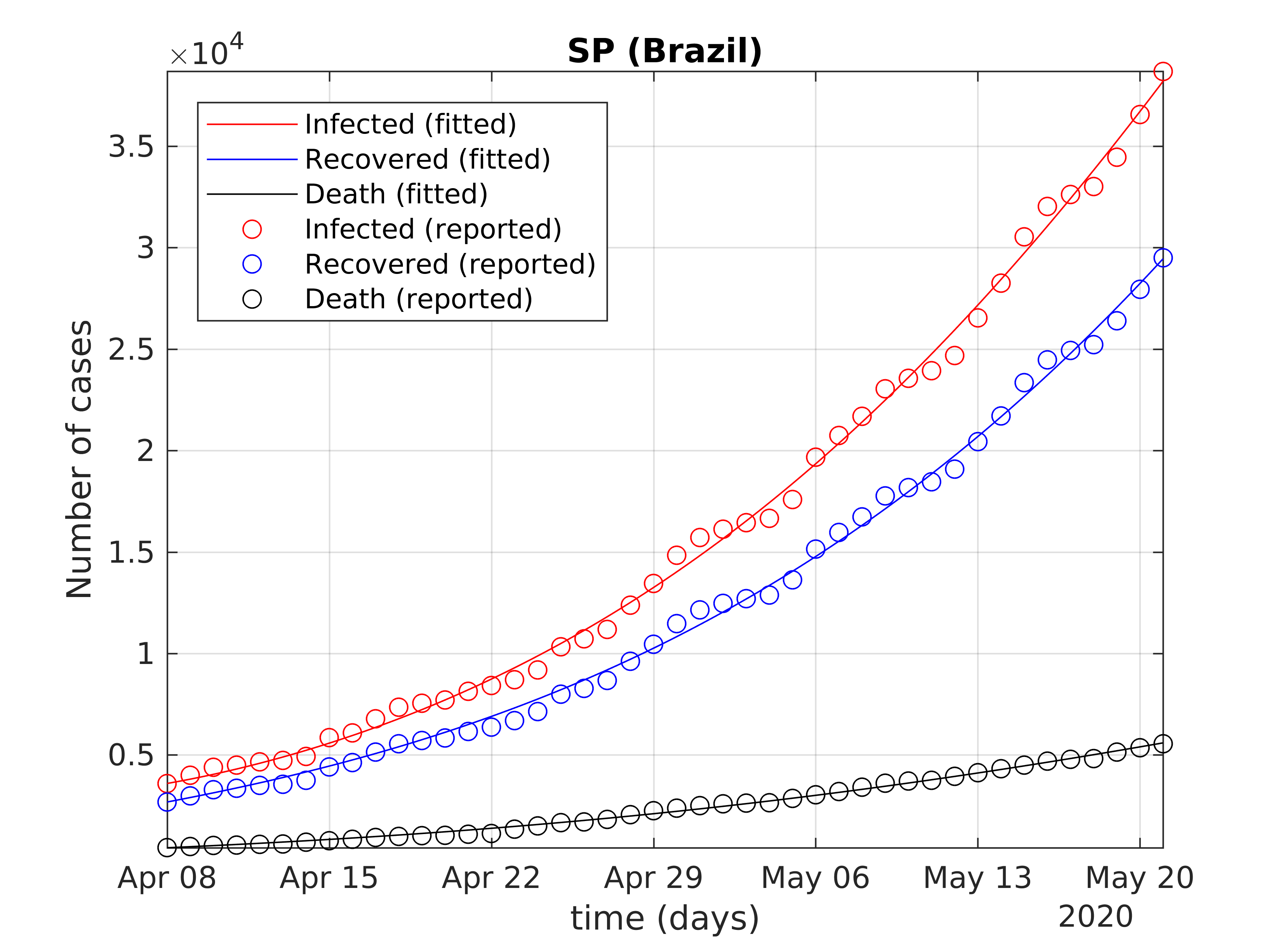}}
	\subfigure[]{\label{Erro_States_Br1}\includegraphics[scale=.5]{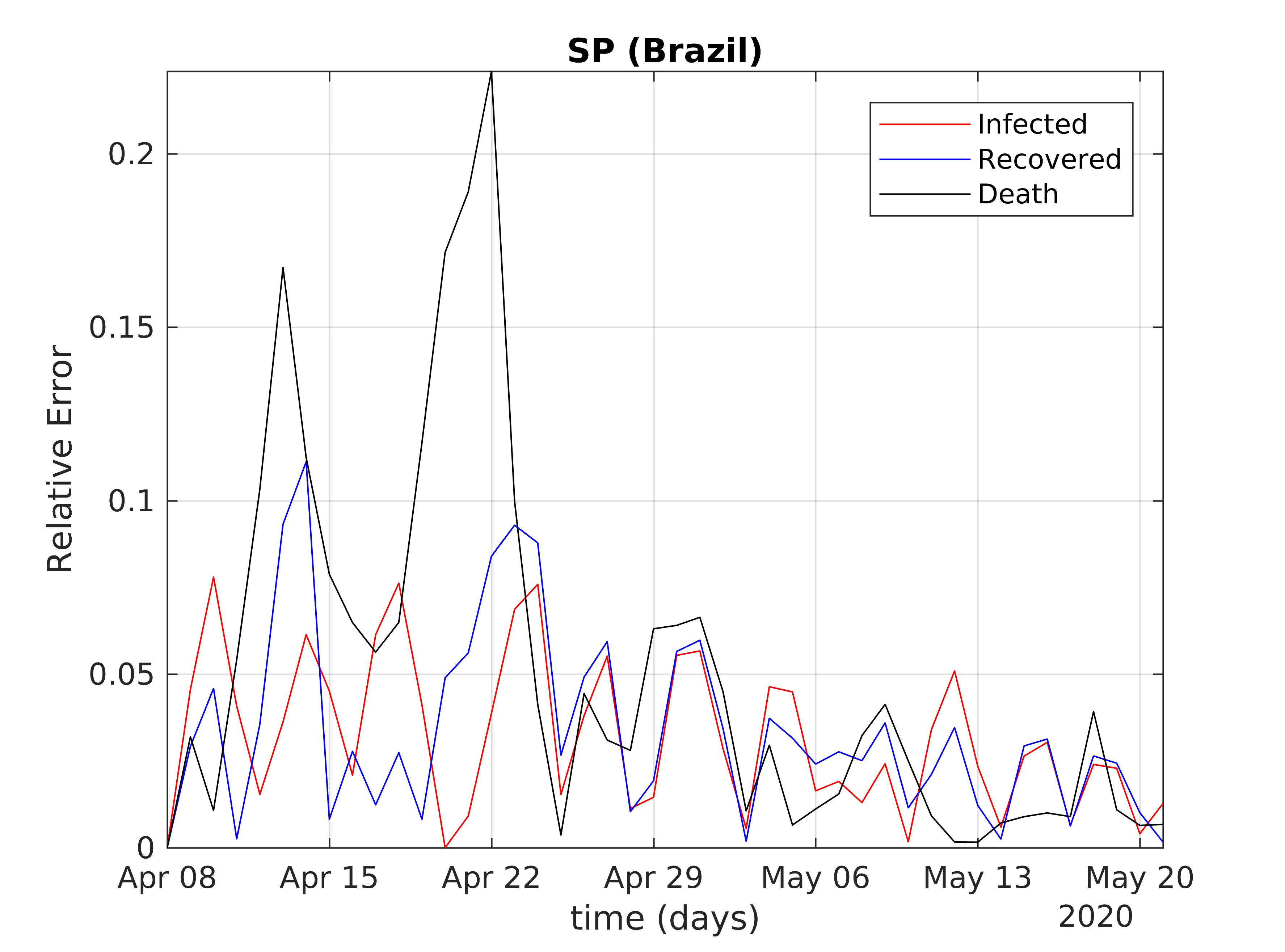}}
	\caption{ (a) Modeling of the differential SEIRDP model from  April, 08 to May 21, 2020. (b) The relative mean errors of projection for Brazil from April 08 to  May 21, 2020 between actual data and SEIRDP model involving cases of infected, recovered and death.}
	\label{fig_SP_Prev}
\end{figure}

São Paulo is the second metropolis in Latin America and the main financial, commercial and corporate center in South America and where the first case of contamination in Brazil was detected. Currently, São Paulo concentrates the largest number of infected with the new Coronavirus.
For São Paulo state, the modeling and relative errors from the same period of Brazil are shown in Figs.~\ref{fig_SP_Prev}. We notice that  the  modeled curves present a good approximation with the collected data. In addition, a large part of the relative errors in infected cases, recovery, and death cases is below $20\%$.

\begin{figure}[!ht]
	\centering
	\subfigure[]{\label{fig_StateBrPrev1}\includegraphics[scale=.5]{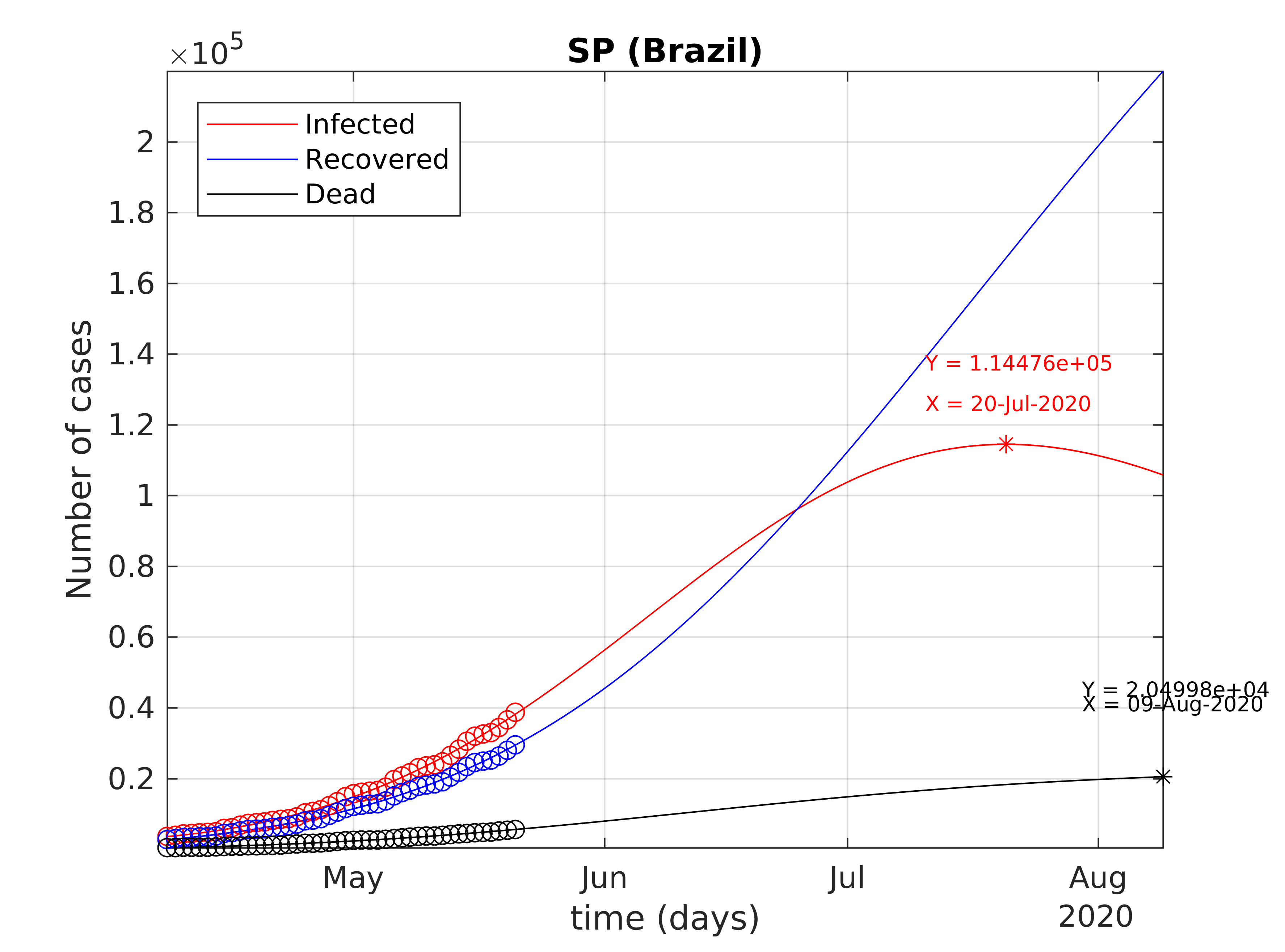}}\\
	\subfigure[]{\label{fig_BrasilConfirmed_D1}\includegraphics[scale=.5]{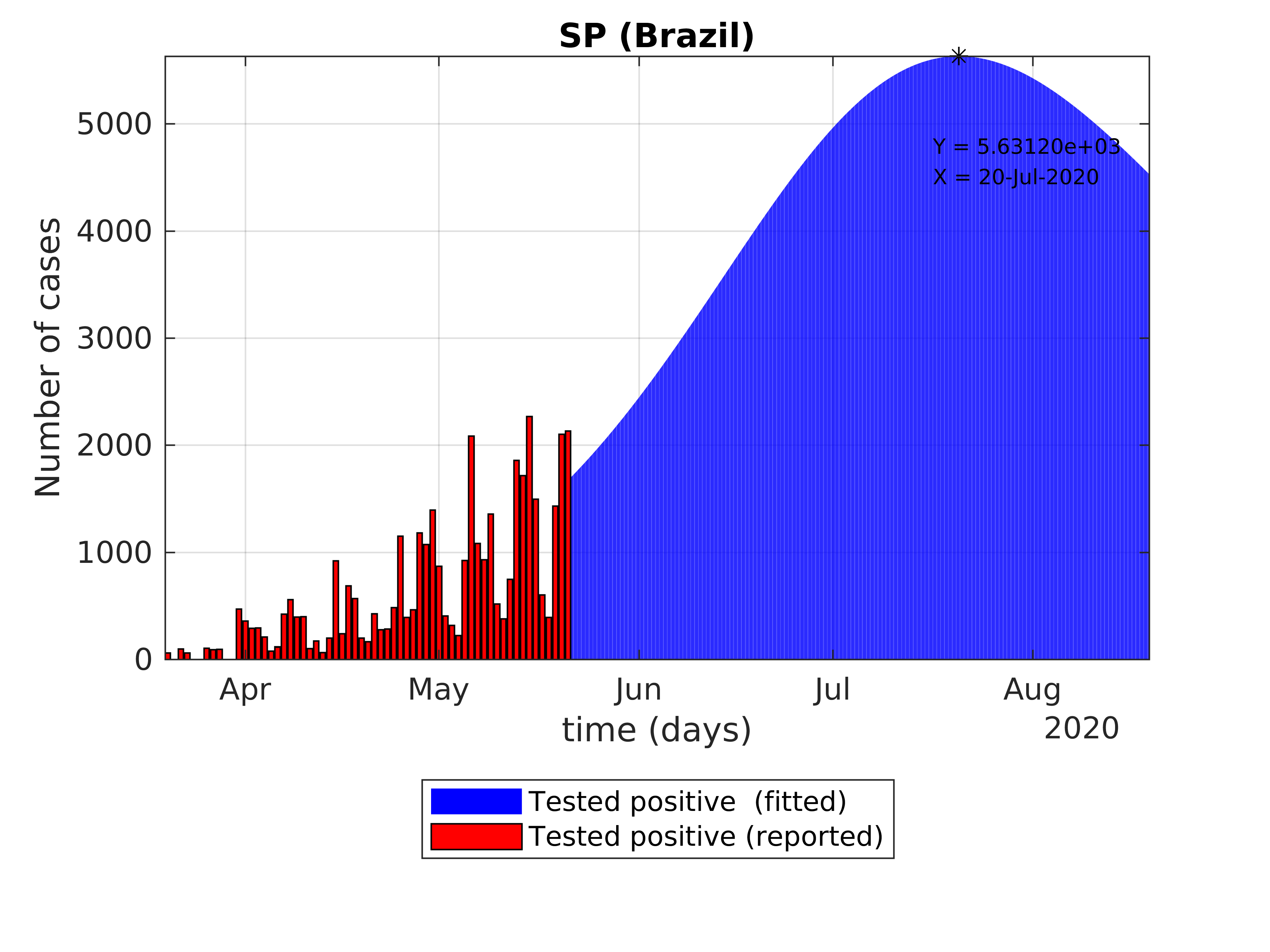}}
	\subfigure[]{\label{fig_BrasilDeath_D1}\includegraphics[scale=.5]{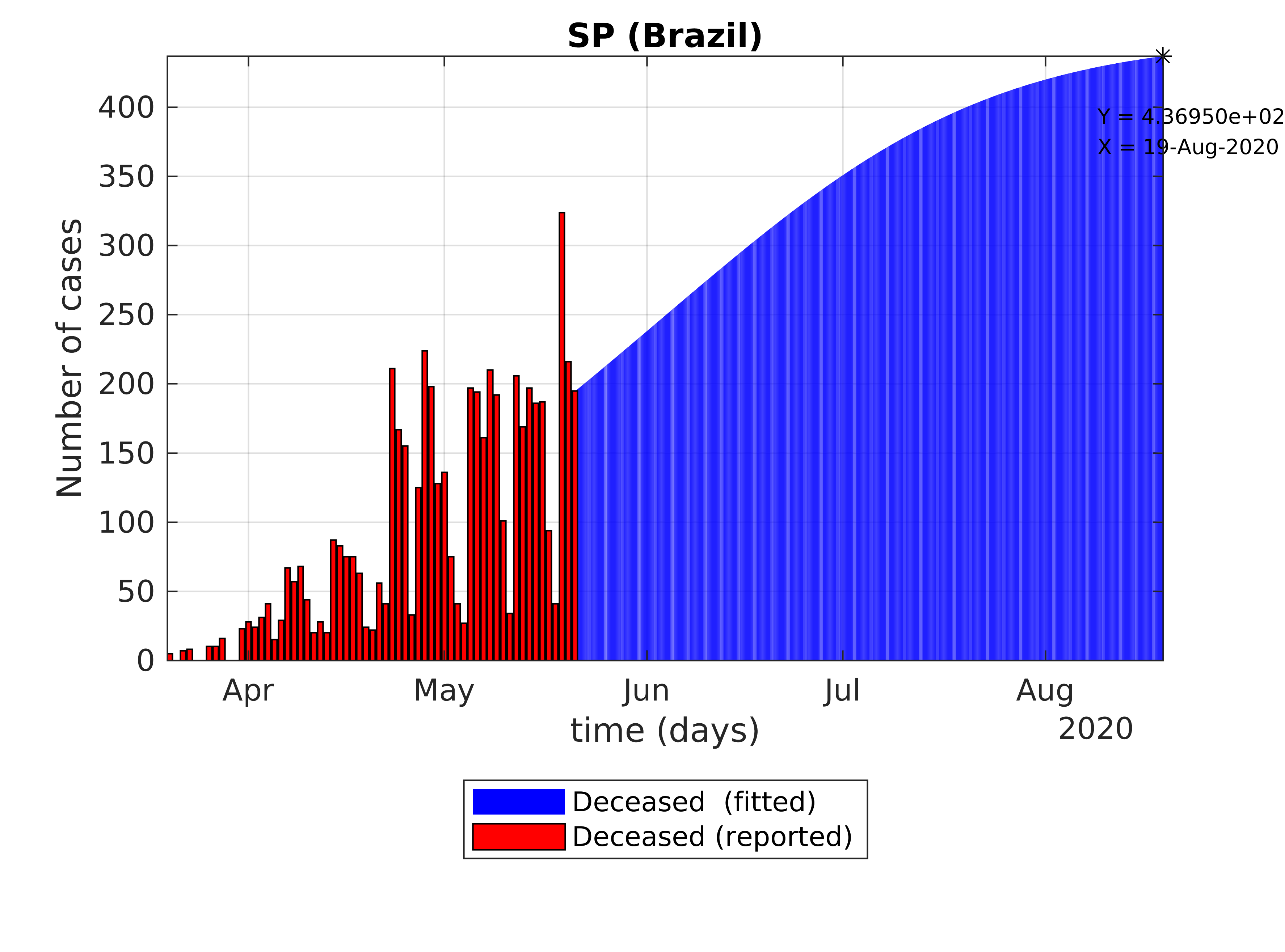}}
	\caption{(a) Predictions of the differential SEIRDP model for S\~ao Paulo state from early April to mid-August, 2020. (b) The daily  of infected cases and (c) deceased measured and projected for S\~ao Paulo  state early April to mid-August, 2020.}
	\label{fig:SEIRDP_SP_BRN}
\end{figure}

In Fig. \ref{fig_StateBrPrev1} we present an estimate of accumulative
infected cases, recovery, and death cases. Projections show that  the currently infected cases will reach
a peak around July 20 close to Brazil,  with about 114K cases.
Fig.~\ref{fig_BrasilConfirmed_D1} and Fig.~\ref{fig_BrasilDeath_D1}  show the daily confirmed
infected cases and the death cases for São Paulo over the mentioned period respectively. Estimates show that the peak will reach in July 20 (about 5K) for daily infected and converging to 436 daily deaths till August 19, 2020. The daily and cumulative peak has shown to be very close to the results obtained in the projection of Brazil.


\subsection{Prediction of Time evolution of COVID-19  for Cear\'a state}\label{Ceara}

\begin{figure}[htb!]
	\subfigure[]{\label{fig_StateBr2}\includegraphics[scale=.5]{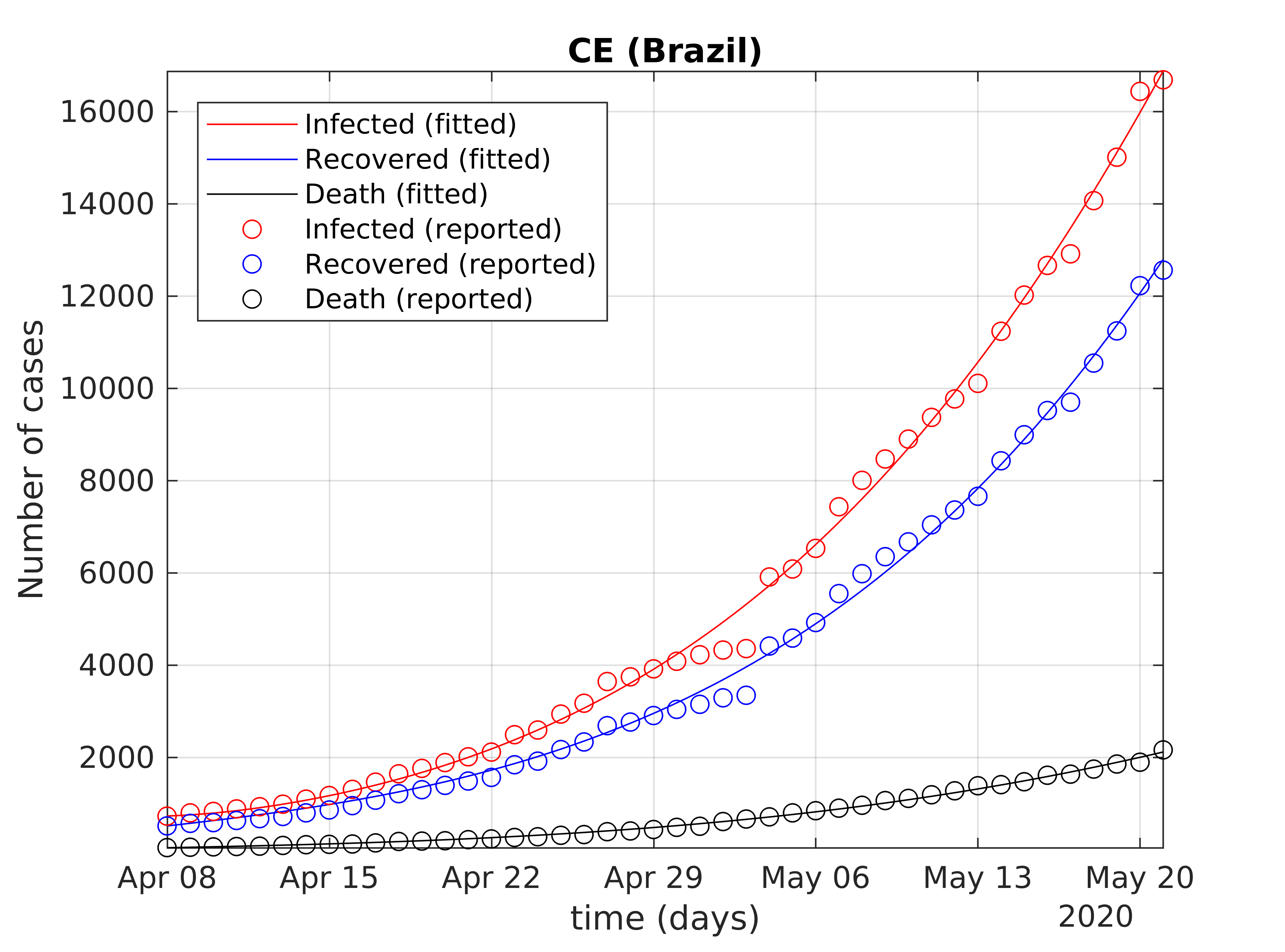}}
	\subfigure[]{\label{Erro_erro_Br2}\includegraphics[scale=.5]{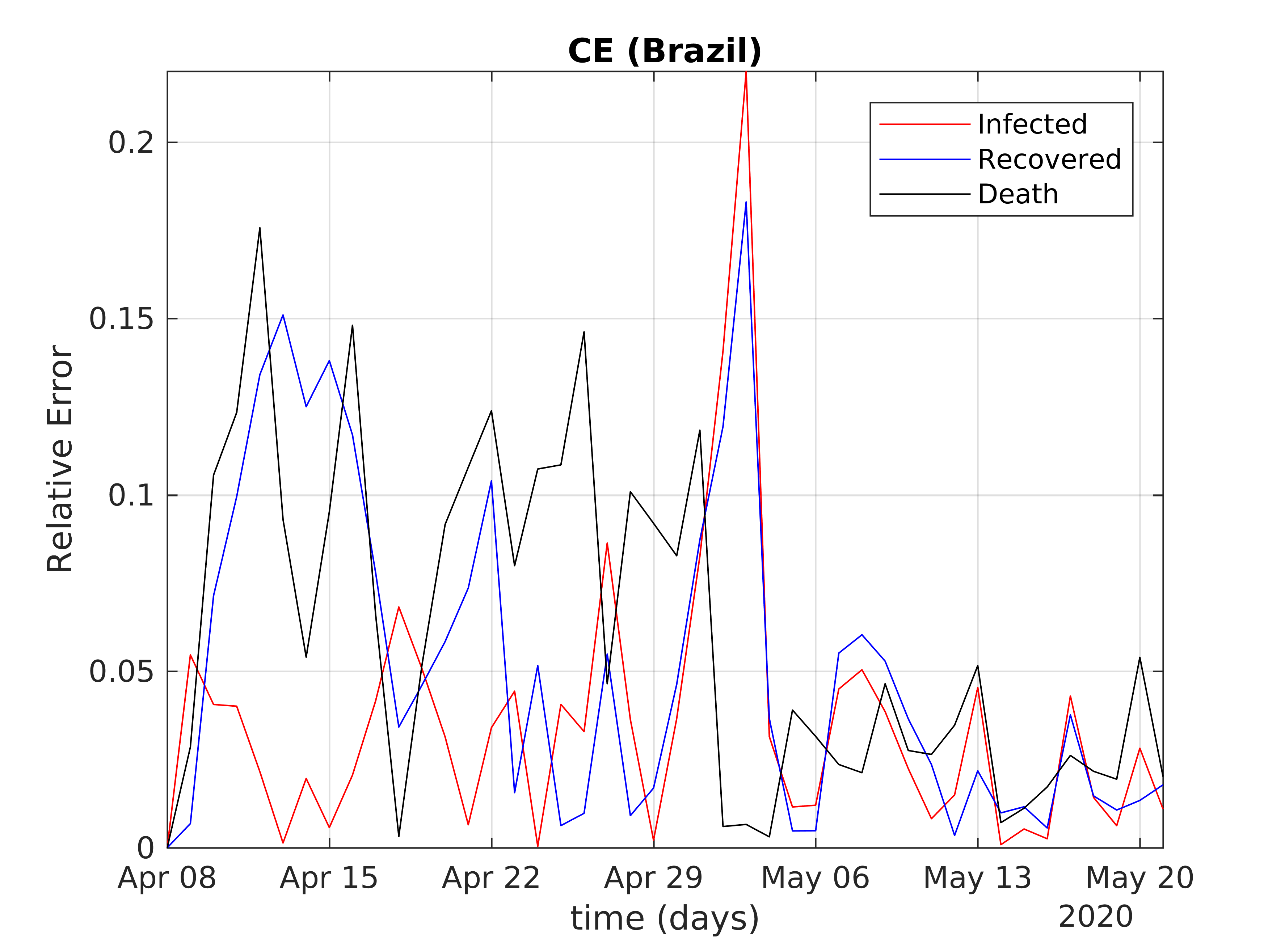}}
	\caption{(a) Predictions of the differential SEIRDP model from  April 08 to May 21, 2020.  (b) The relative mean errors of projection for Ceará state from  April 08 to  May 21, 2020 between actual data and SEIRDP model involving cases of infected, recovered and death. }
	\label{fig_BR_CE}
\end{figure}

Another important study presented is on the state of Ceará, which recently showed a great increase in the number of cases.
For Ceará state, the modeling  and relative errors  from the same period of Brazil are shown in Figs.~\ref{fig_BR_CE}. Looking at the curves modeled for Ceará state, we see that the  error curves of infected, recovered and deceased cases also are  oscillating around less than 20\% in the period from early April to mid-May. Therefore, it is possible to infer that the SEIRDP model provides a good approximation of the results with respect to the collected data.

\begin{figure}[!ht]
	\centering
	\subfigure[]{\label{fig_StateBrPrev21}\includegraphics[scale=.5]{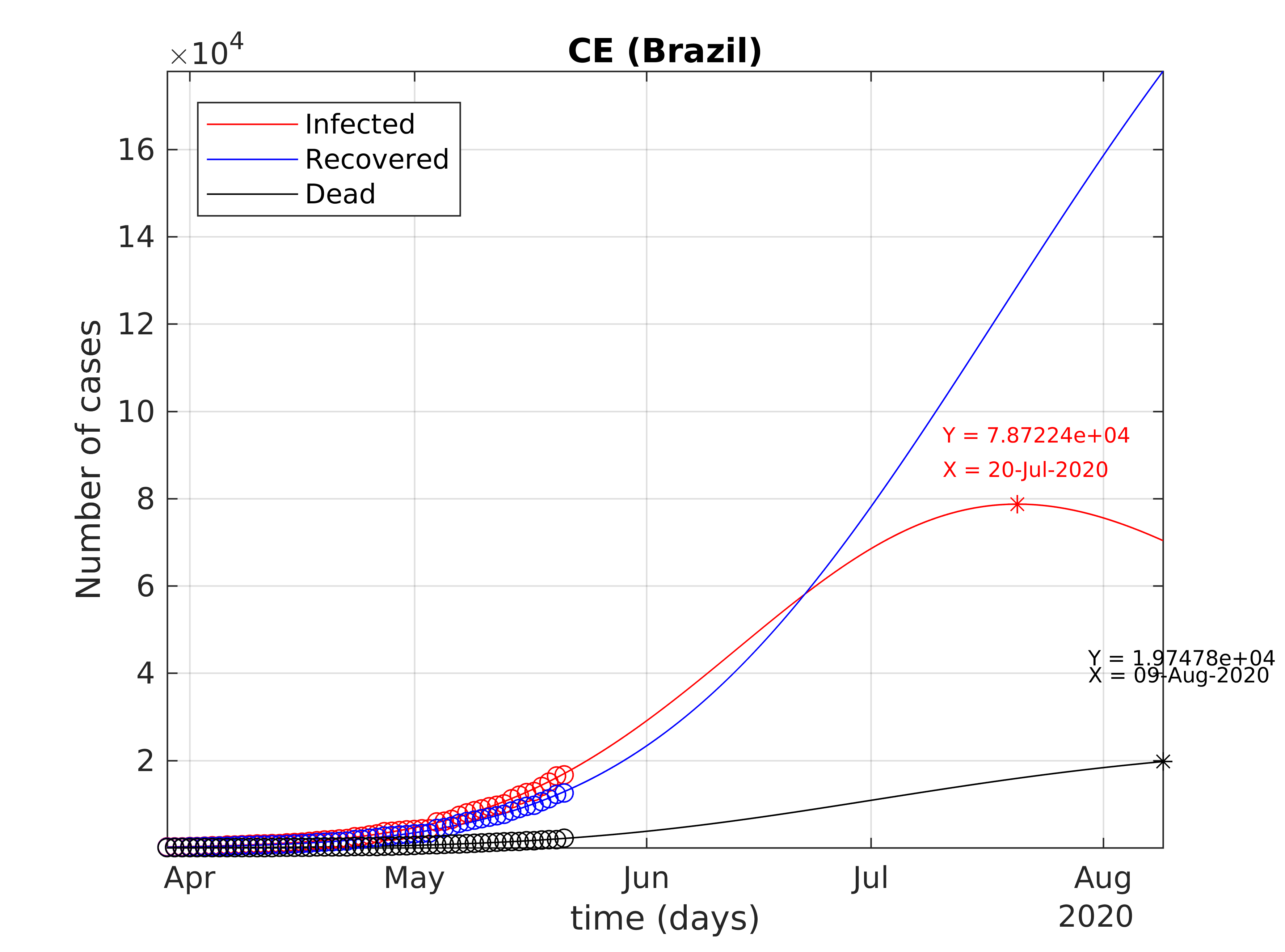}}\\
	\subfigure[]{\label{fig_BrasilConfirmed_D21}\includegraphics[scale=.5]{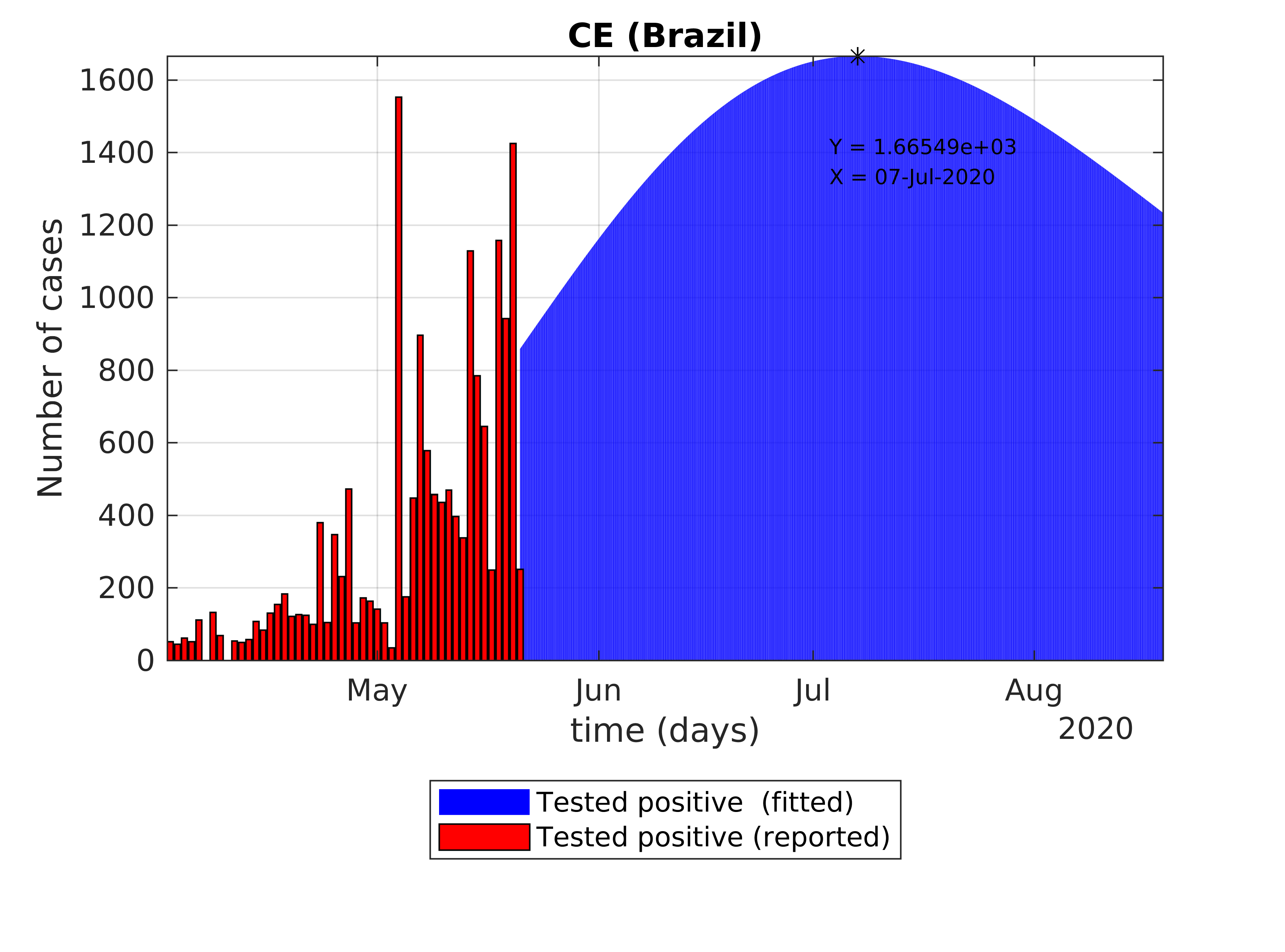}}
	\subfigure[]{\label{fig_BrasilDeath_D21}\includegraphics[scale=.5]{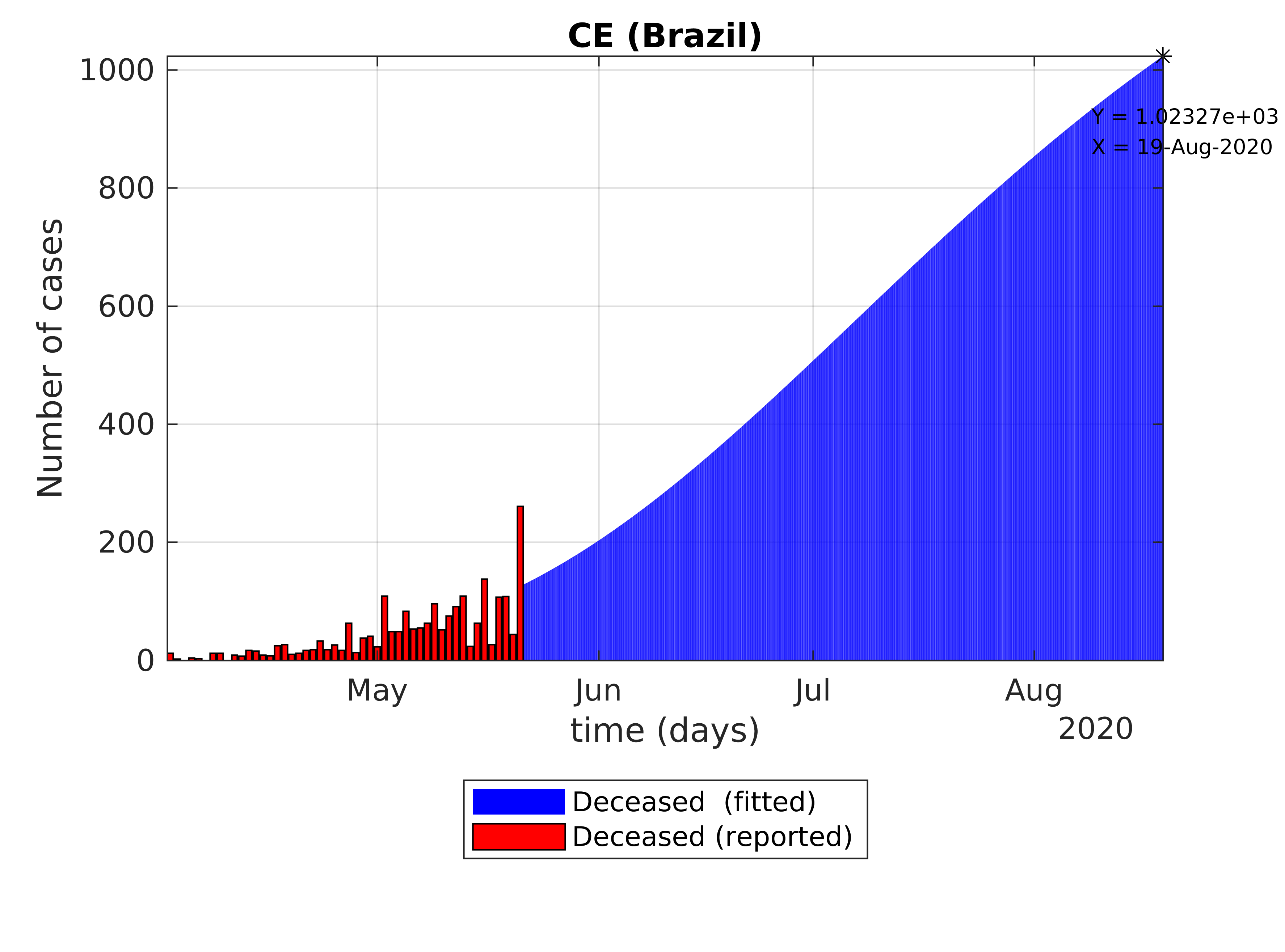}}
	\caption{(a) Predictions of the differential SEIRDP model for Ceará state from	early April to early August, 2020. (b) The daily  of infected cases and (c) deceased measured and projected for Ceará state early April to early August, 2020.}
	\label{fig:SEIRDP_CE_BRN}
\end{figure}

In Fig. \ref{fig_StateBrPrev21} we provide an estimate of  accumulative
infected cases, recovery and death cases. Projections show that  the currently infected cases will reach
a peak  around July 20 with about 80K cases.
Fig.~\ref{fig_BrasilConfirmed_D21} and Fig.~\ref{fig_BrasilDeath_D21} present the daily confirmed
infected cases and  the death cases  for Ceará over mentioned period, respectively. By the estimates, we can conjecture that the peak will reach in July 07 (about 1665) for daily infected, converging to 1000 daily deaths till August 19, 2020.

\subsection{Prediction of Time evolution of COVID-19 for Rio de Janeiro state}

\begin{figure}[htb]
	\subfigure[]{\label{fig_StateBrRJ}\includegraphics[scale=.5]{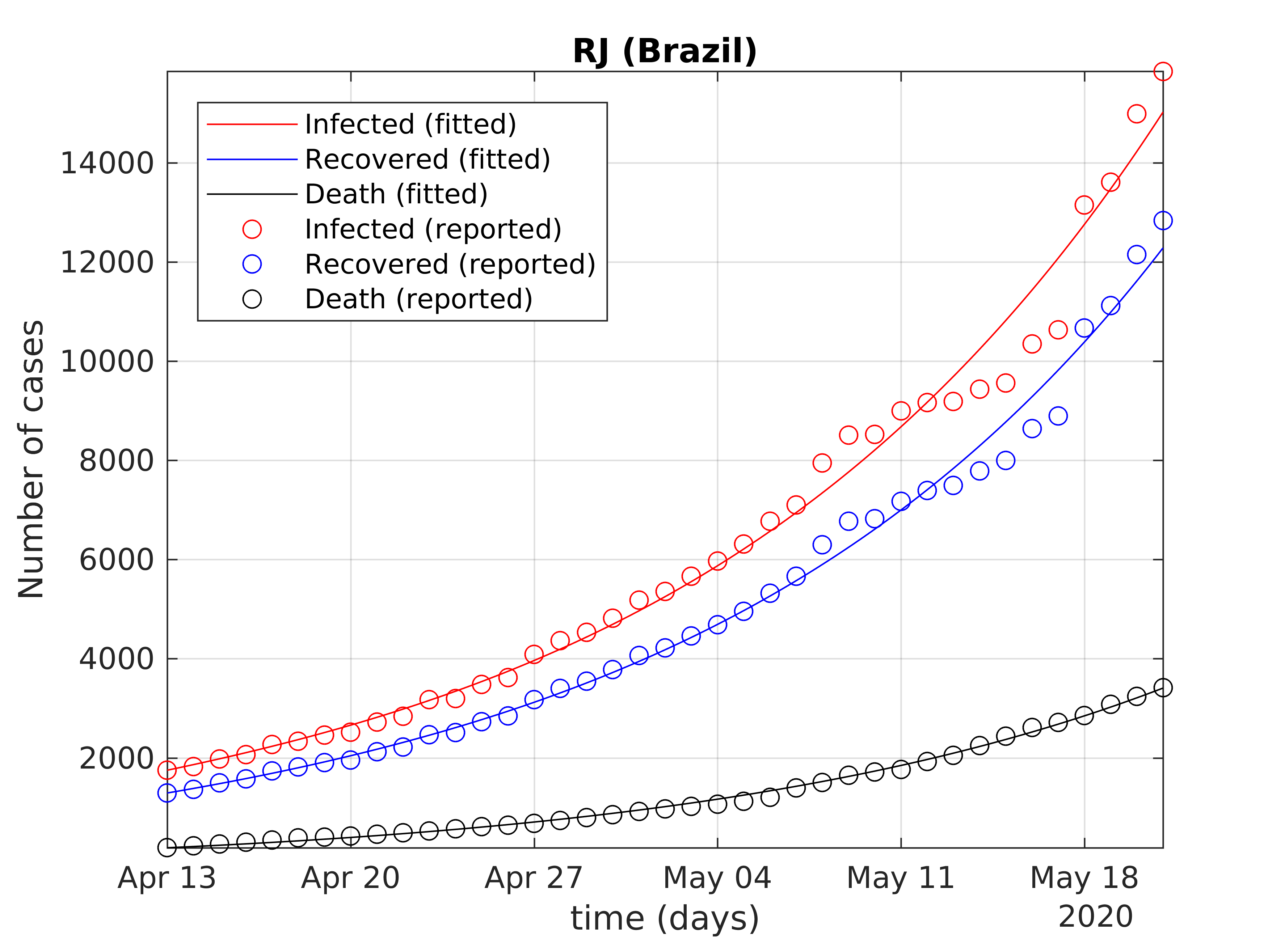}}
	\subfigure[]{\label{Erro_erro_BrRJ}\includegraphics[scale=.5]{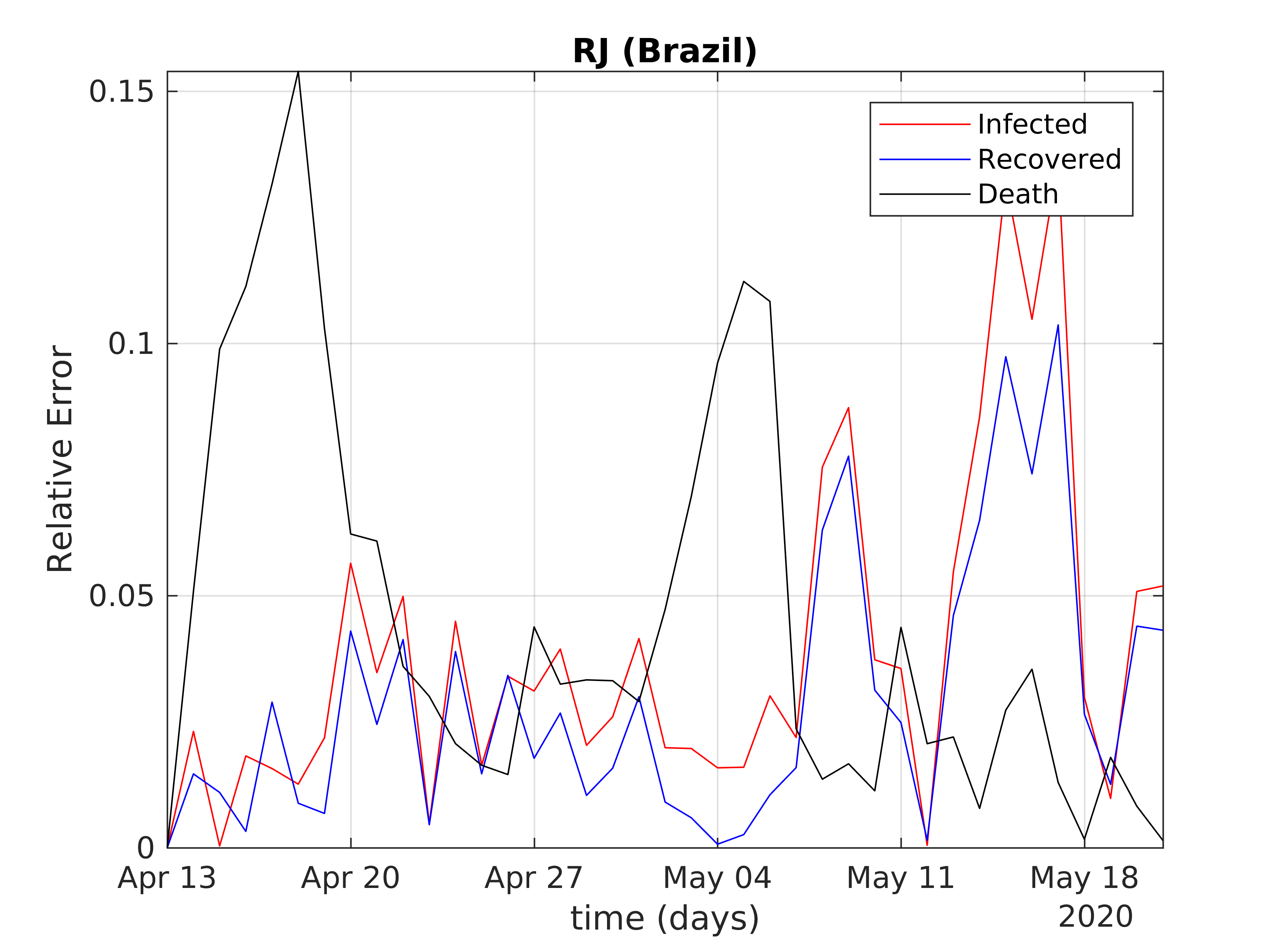}}
	\caption{(a) Predictions of the differential SEIRDP model from  Match 20 to May 17, 2020.  (b) The relative mean errors of projection for Rio de Janeiro State from  April 08 to  May 17, 2020 between actual data and SEIRDP model involving cases of infected, recovered and death. }
	\label{fig_BR_RJ}
\end{figure}

We also present a more detailed study of the second largest state in population of Brazil and the second state to record cases of contaminated by the new Coronavirus which is Rio de Janeiro.
The modeling and projection results from the same
period of Brazil are shown in Fig.~\ref{fig_StateBrRJ} for infected, recovered and deaths.  The results from April to middle May bring consistent response.  The respective relative errors are shown in Fig. \ref{Erro_erro_BrRJ}).

\begin{figure}[!ht]
	\centering
	\subfigure[]{\label{fig_StateBrPrevRJ}\includegraphics[scale=.5]{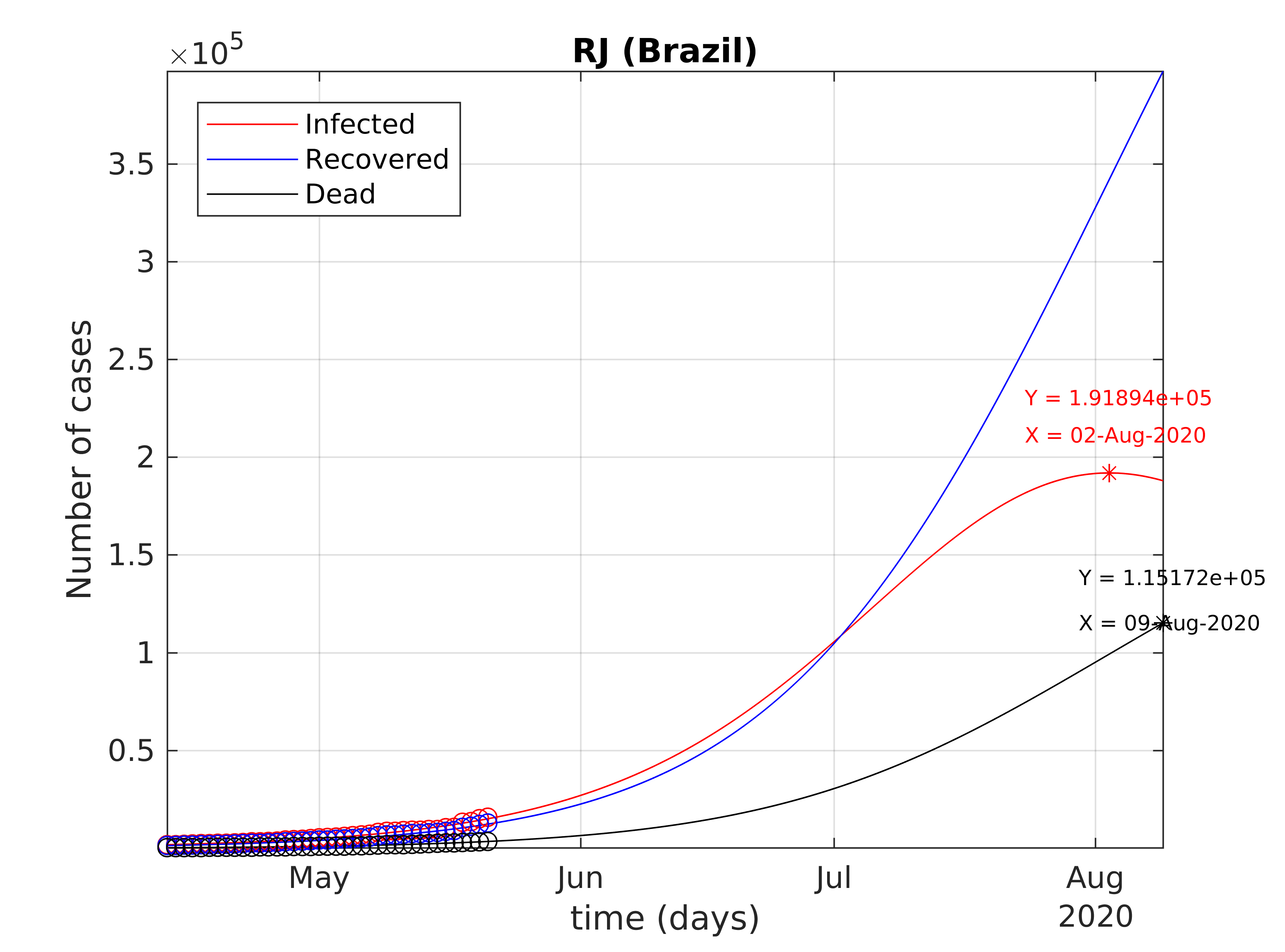}}\\
	\subfigure[]{\label{fig_BrasilConfirmed_DRJ}\includegraphics[scale=.5]{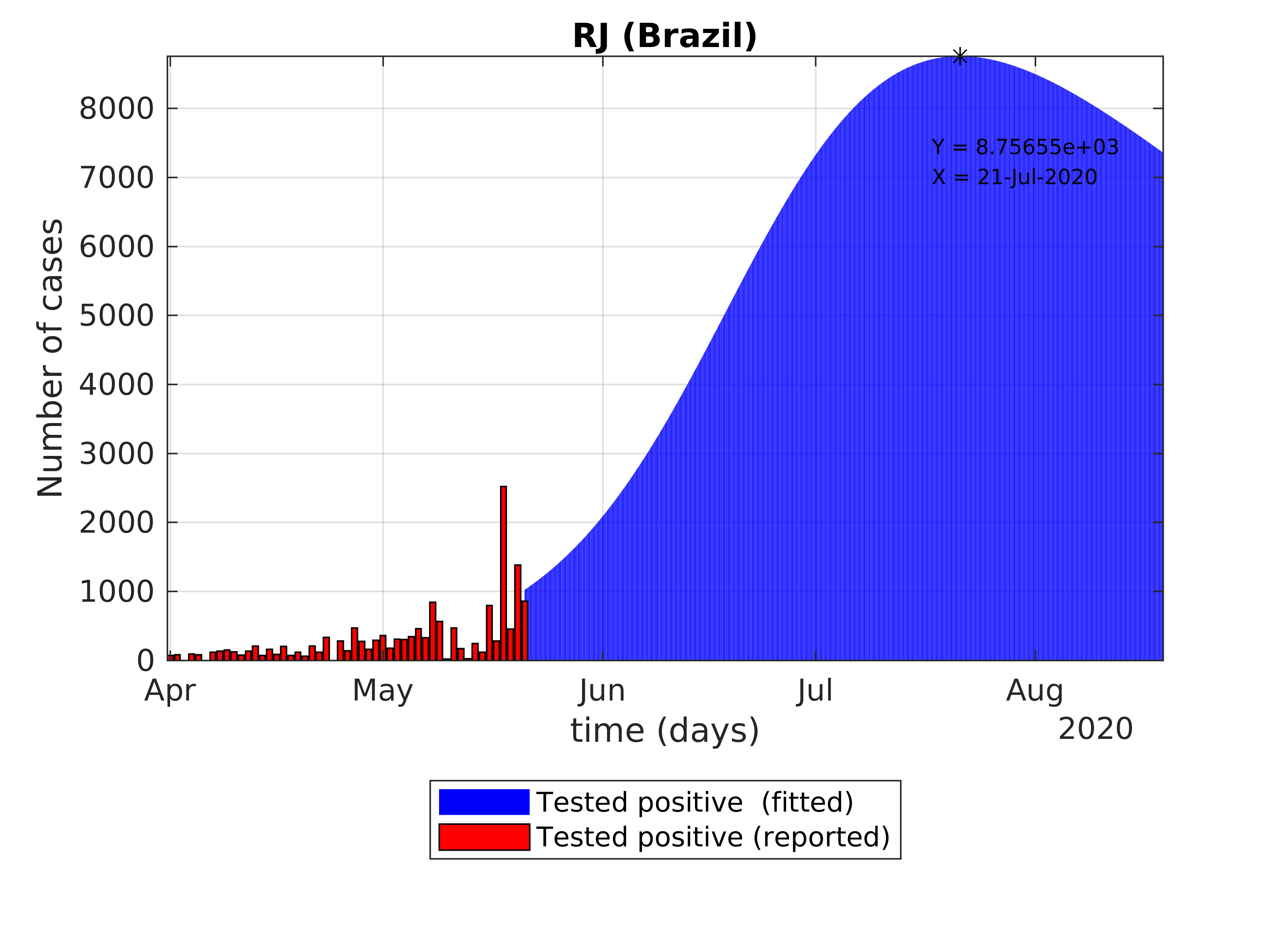}}
	\subfigure[]{\label{fig_BrasilDeath_DRJ}\includegraphics[scale=.5]{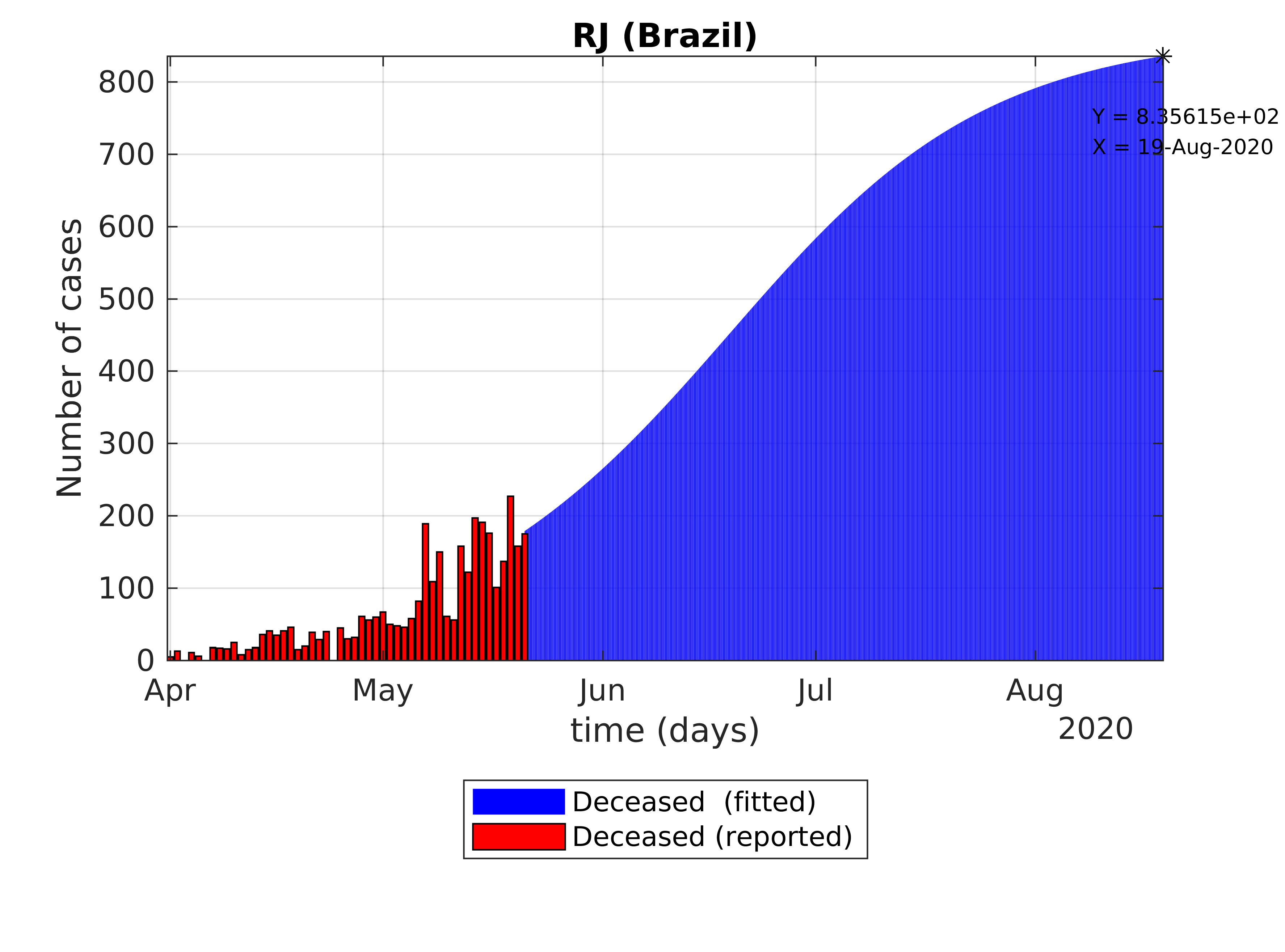}}
	\caption{(a) Predictions of the differential SEIRDP model for Rio de Janeiro state from	early April to early August, 2020. (b) The daily  of infected cases and (c) deceased measured and projected for Rio de Janeiro state early April to  mid-August, 2020.}
	\label{fig:SEIRDP_RJ_BRN}
\end{figure}

According to Fig.~\ref{fig_StateBrPrevRJ}, the state of Rio de Janeiro projects that infected cases  will reach to the peak around  August 02  with about 191K cases, which is close to period of  the peak estimation of Brazil. This plot also suggests that the number of deaths will continue with the same growth trend in the cumulative total if the population is not subject to new rules of protection.

Fig.~\ref{fig_BrasilConfirmed_DRJ} and
Fig.~\ref{fig_BrasilDeath_DRJ}  show the daily confirmed
infected cases, the death cases and the total accumulative infected
cases for Rio de Janeiro state over mentioned period, respectively.  Rio de Janeiro
state will reach the peak around July 21 for both the daily infected
increase (about 9K per day). But for daily death cases, it will reach
the peak of 800 per day around August 19.

\subsection{Prediction of time evolution of COVID-19 for other states}

In this study, a time-dependent SEIRDP model  was used to analyze the evolution of currently active, recovered and deaths
cases in states in Brazil whose infected toll exceeds 8.000 cases actually. The states Espírito Santo, Bahia, Amazonas, Pará, Pernambuco and Maranhão marked together with São Paulo, Ceará, and Rio de Janeiro have more than 90 percent of the total fatalities in the country. As in previous experiments, theses projections take into account mitigation and self-protective measures implemented by the government and the population, in the periods given previously.

\begin{figure}[!ht]
	\centering
	\subfigure[]{\label{fig_StateBrPrev2}\includegraphics[scale=.39]{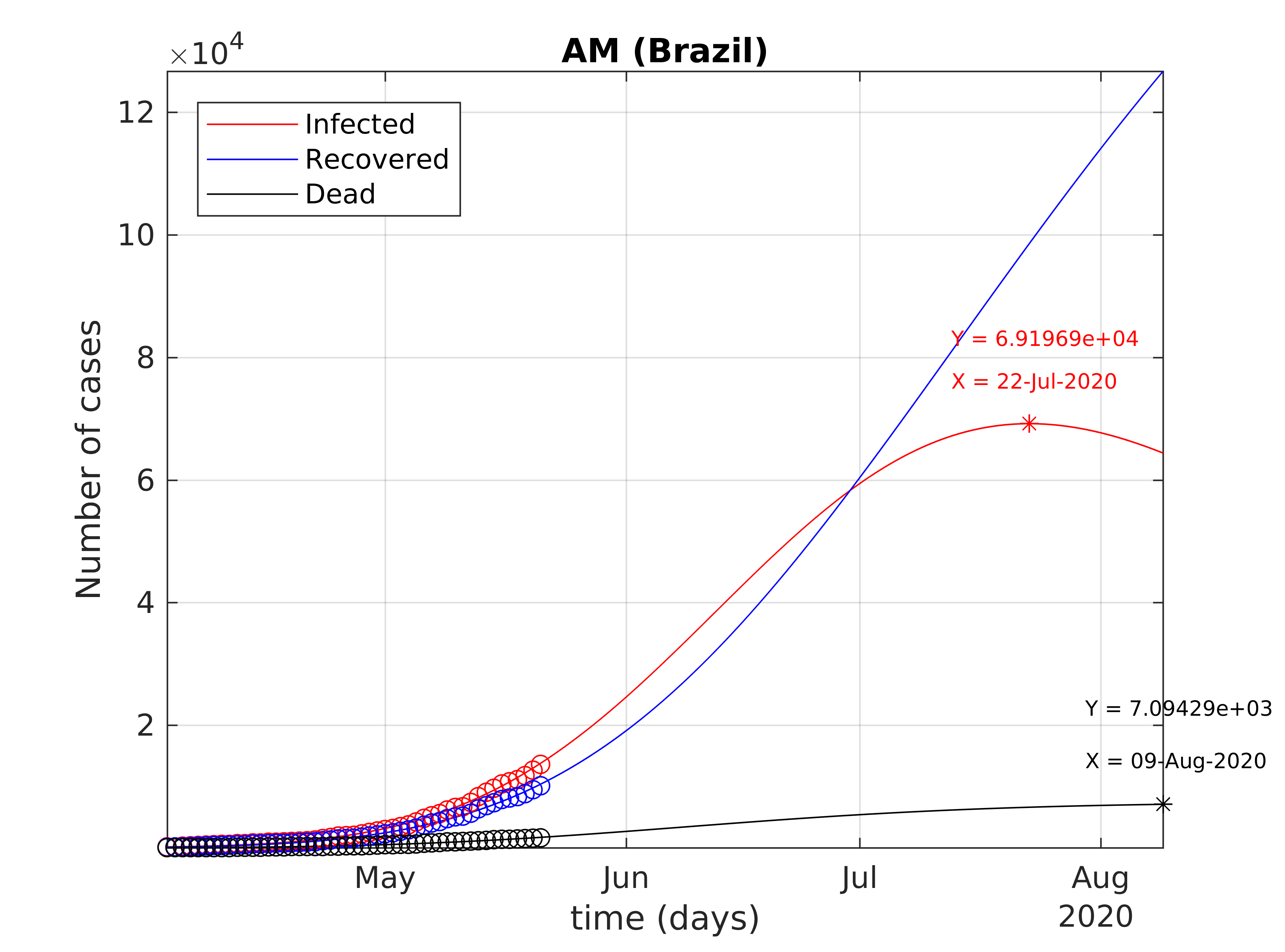}}
	\subfigure[]{\label{fig_StateBrPrev2}\includegraphics[scale=.5]{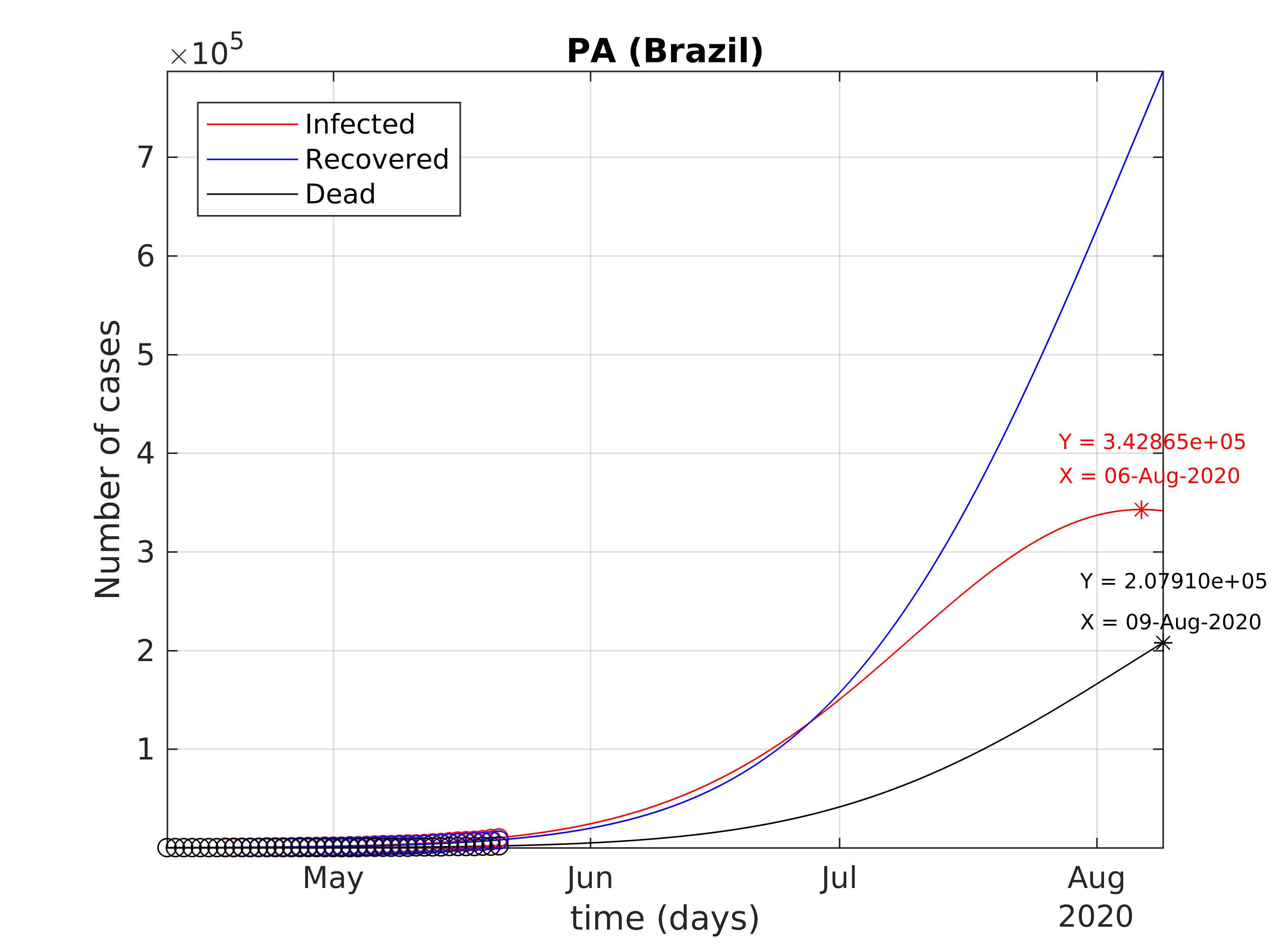}}
	\subfigure[]{\label{fig_StateBrPrev2}\includegraphics[scale=.5]{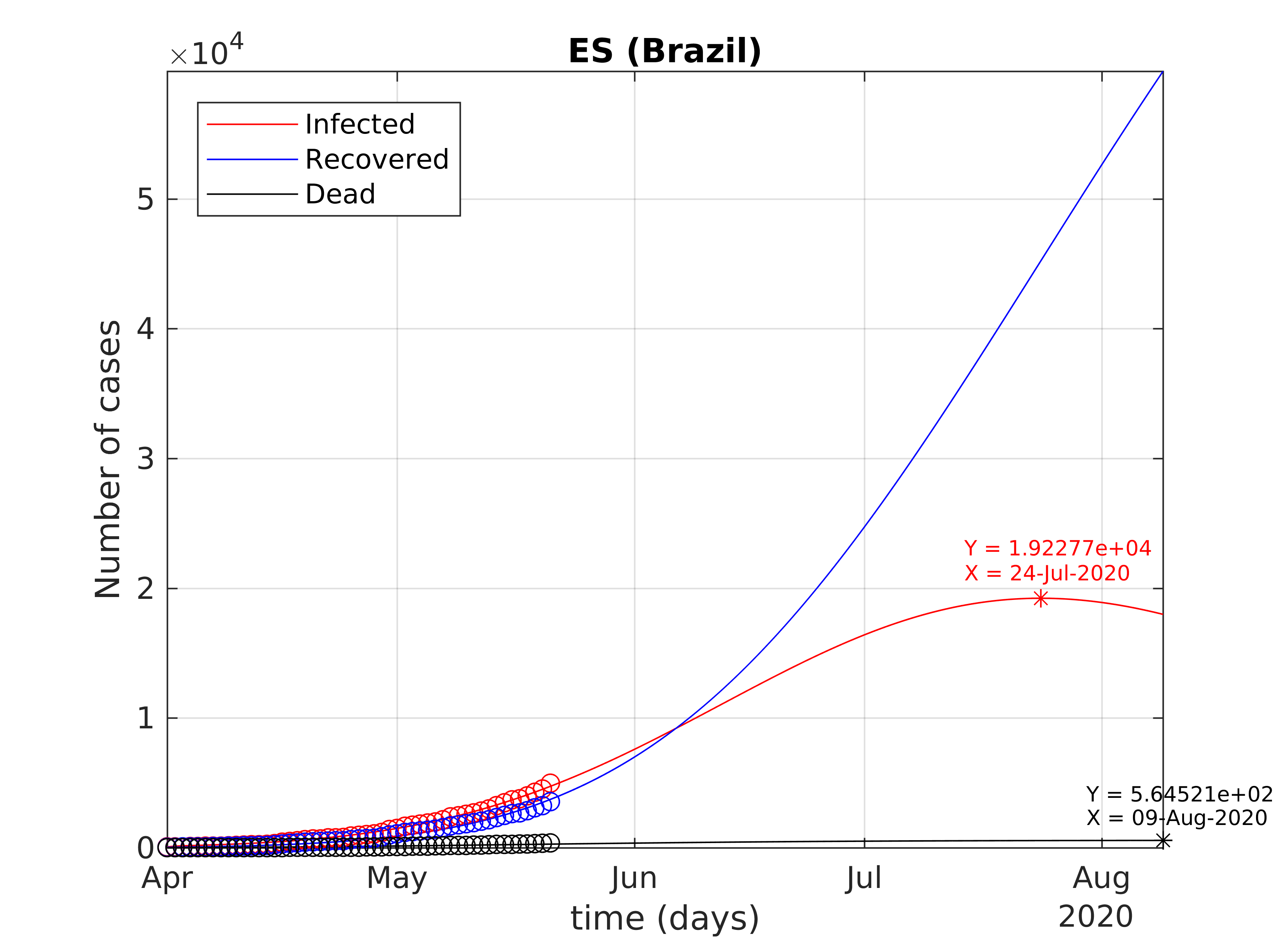}}
	\subfigure[]{\label{fig_StateBrPrev2}\includegraphics[scale=.5]{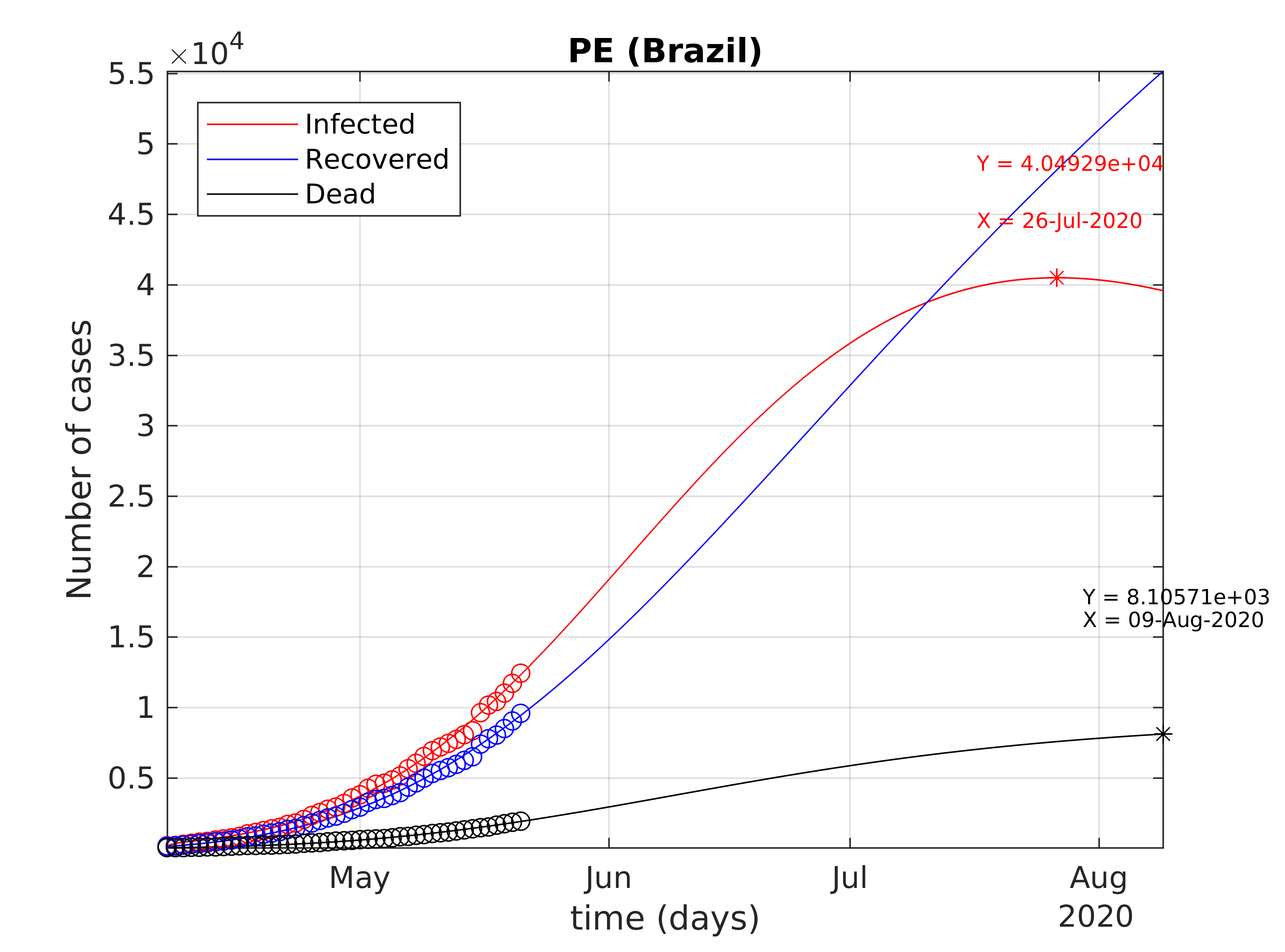}}
	\subfigure[]{\label{fig_StateBrPrev2}\includegraphics[scale=.5]{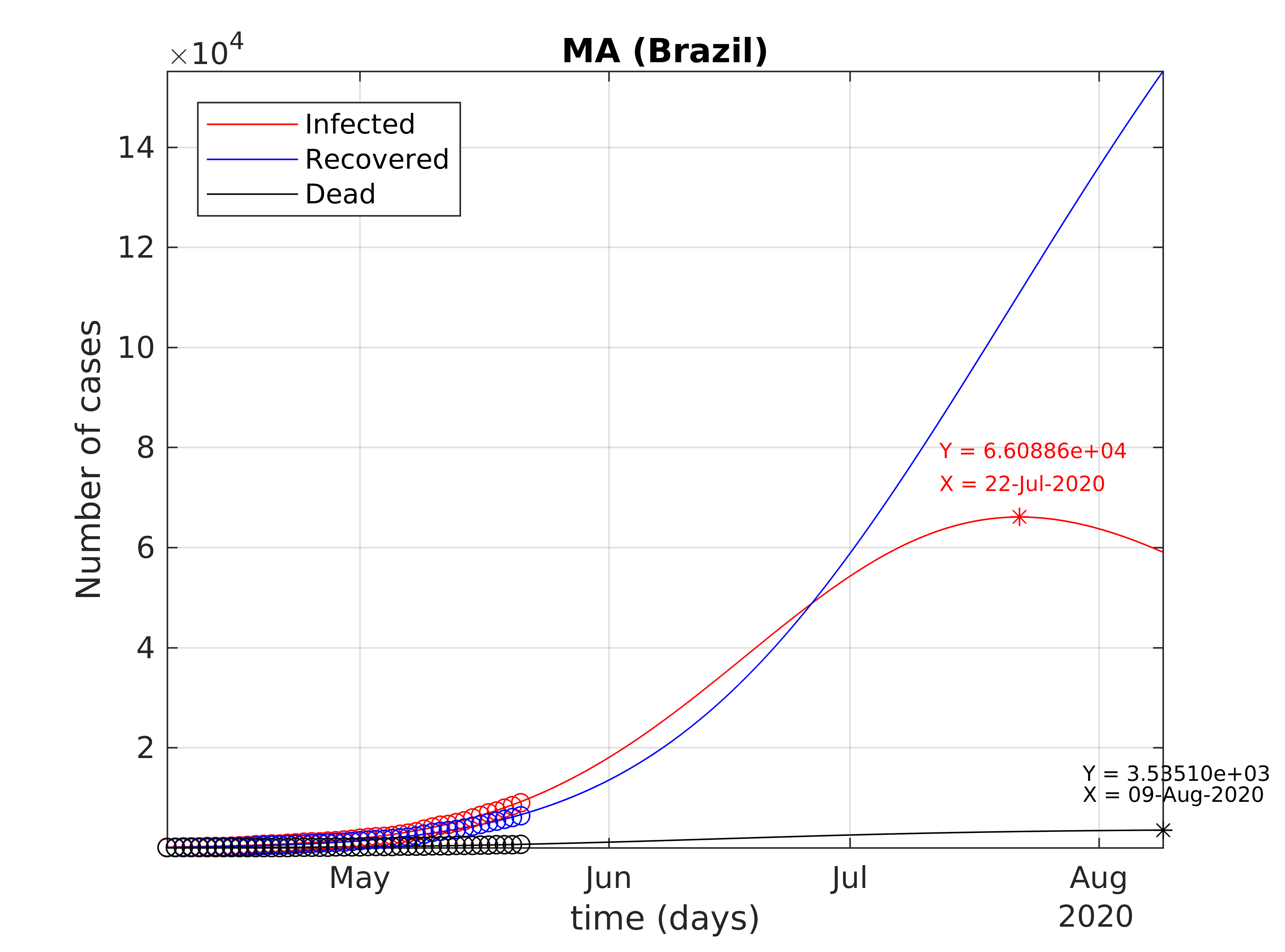}}
	\subfigure[]{\label{fig_StateBrPrev2}\includegraphics[scale=.5]{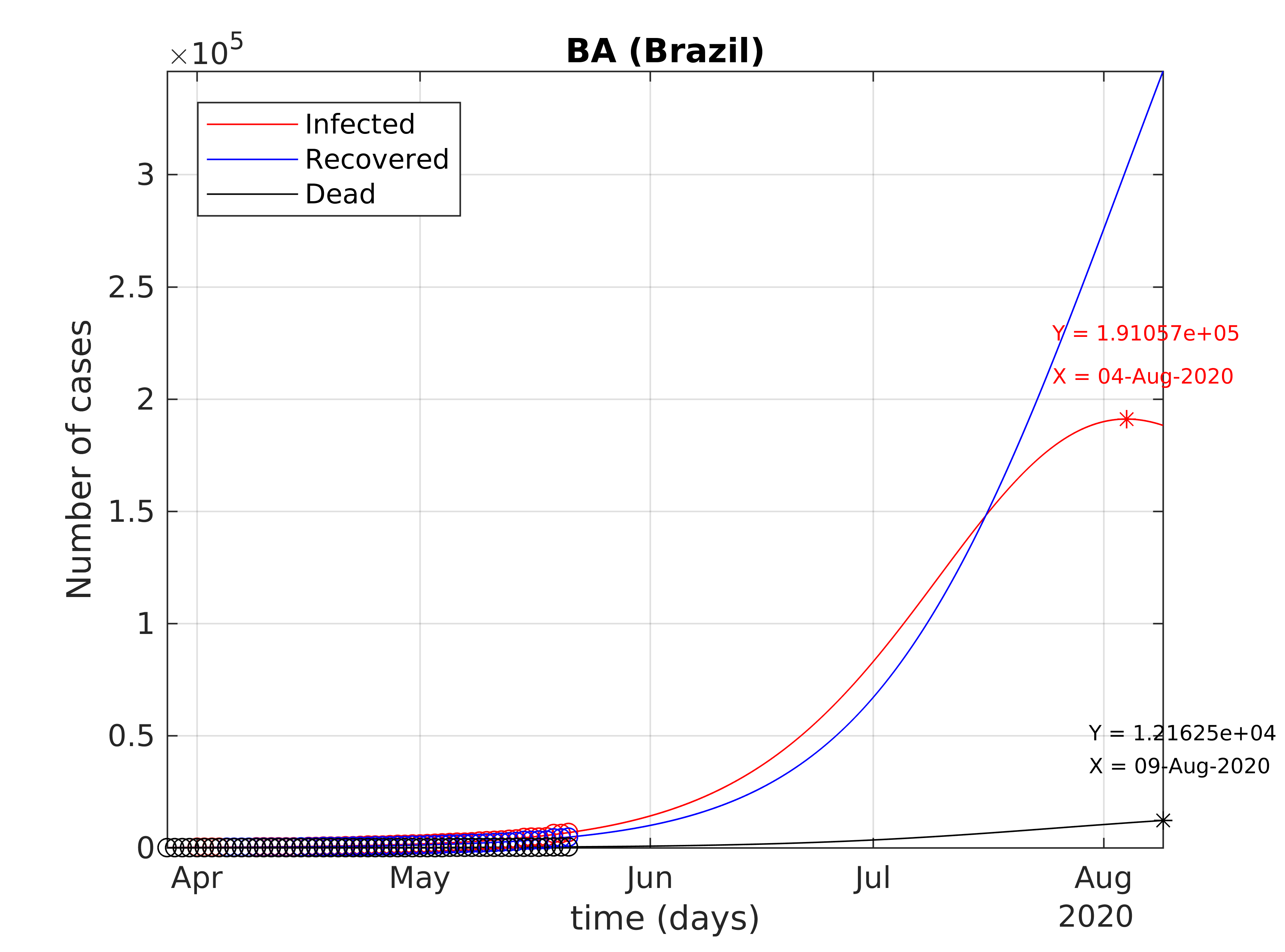}}
	\caption{ Predictions of the differential SEIRDP model for Amazonas (AM), Pará (PA), Espirito Santos (ES), Pernambuco (PE), Maranhão (MA)   and Bahia (BA) from	early April to early August, 2020.}
	\label{fig:SEIRDP_OT_BRN}
\end{figure}

Fig.\ref{fig:SEIRDP_OT_BRN} shows the prediction of infected, recovered, and death from April to early August. According to this figure, the state of Pernambuco is close to reaching the peak of patients infected by COVID-19 in mid-June, giving the authorities time to plan control and mitigation measures during this period. On the other hand, the number of deaths maintains a projected growth in all states considered in this experiment.

\section{Discussions and Conclusions}

We consider an extension of the classical SEIR model  known as SEIRDP with the aim of modeling COVID-19 epidemics and introduce an estimation algorithm for Brazil. These parameters are found by applying the least-squares estimate in order to fit the SEIR differential equations to the daily and cumulative cases observed by symptoms onset. The results revealed that the  model \eqref{md:SEIR} provides a good approximation with respect to the collected data, since a large part of the relative errors in infected cases, recovery, and death is below $ 20\%$.

\begin{figure}[!ht]
	\centering
	\subfigure[]{\label{fig_StateBrAjustPartial}\includegraphics[scale=.5]{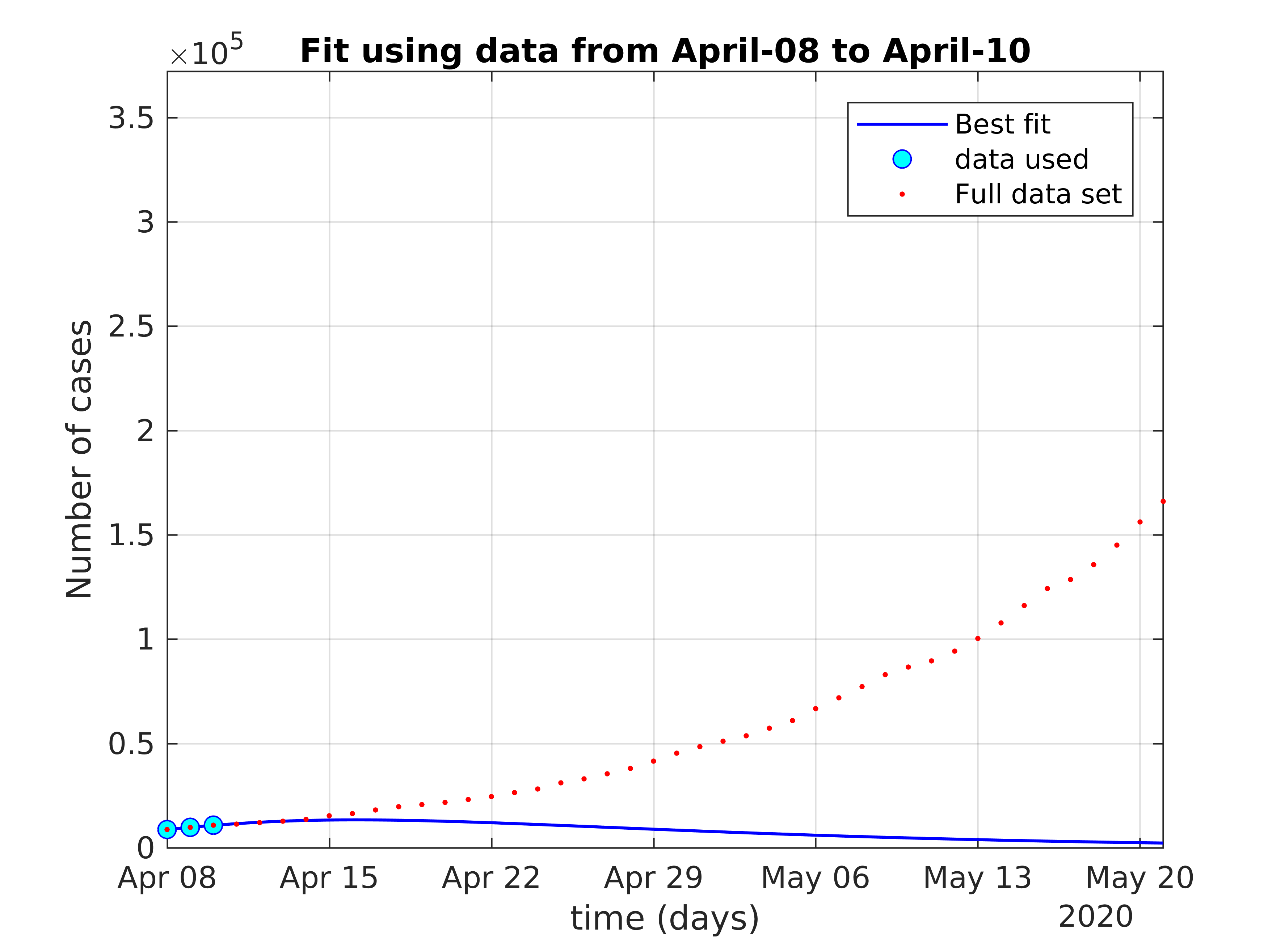}}
	\subfigure[]{\label{fig_StateBrAjusttotal_2}\includegraphics[scale=.5]{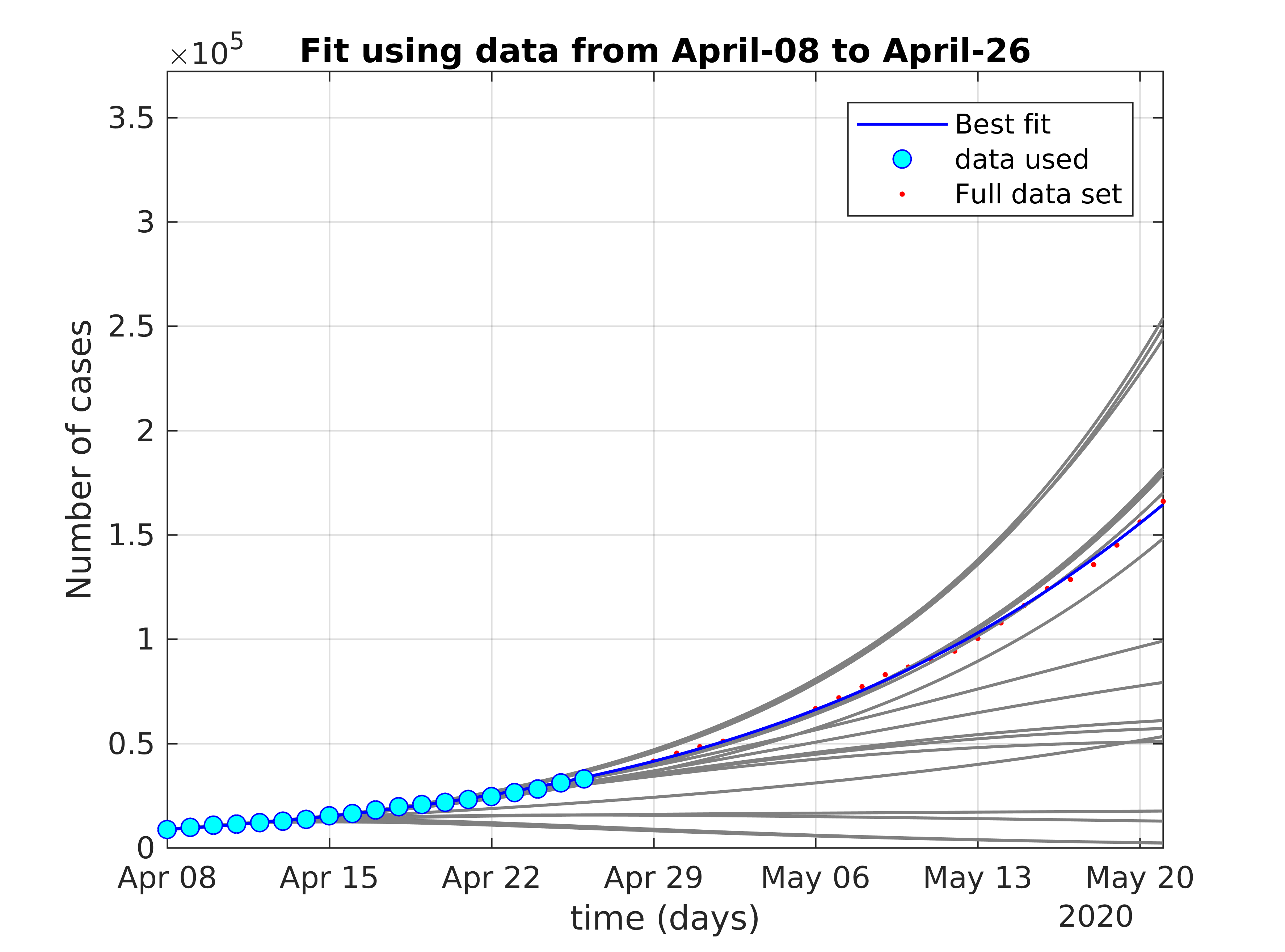}}
	\caption{ (a) Prediction of the differential SEIRDP model for Brazil   from	 April-08 to April-10.  (b) Several predictions of the differential SEIRDP model for Brazil   from	 April-8 to April-26.}
	\label{fig:SEIRDP_BR_BRN}
\end{figure}

It should be noted that we need a reasonable set of data so that the estimates are more reliable.
Fig. \ref{fig:SEIRDP_BR_BRN}  illustrates how bad can be the prediction if the data set is too small. The case of Brazil is used in the following display. Note that the discrepancies are significant when few points are used to measure the prediction of results, but these discrepancies tend to reduce as the number of points increases.

According to the estimates obtained through the SEIRDP model, the states of São Paulo, Ceará, Amazonas, Maranhão, Espírito Santo, Pará and Bahia will reach to the peak of patients infected in mid-July, while Rio de Janeiro in early august. In addition, the daily number of deaths projects growth until August 19, 2020.

It is worth mentioning that the first cases were diagnosed in the country, suspected cases, and those who had contacts with confirmed cases were tested. Due to the large scale of contamination and low testing capacity, the Ministry of Health started to recommend since mid-March that in places with community transmission only serious cases should be tested. Consequently, many cases of coronavirus may have been underreported, underestimating the results of the model. Another intrinsic problem raised by this study was the delay between the occurrences of deaths and their recording in official data released by the Ministry of Health. Unfortunately, in some states, much information arrived late, this can compromise the projection results. Despite these problems, we hope that the estimates obtained will minimize the influence of Coronavirus in Brazil,  especially in large urban centers whose probability of accumulating serious cases in the short term is high.

%

\end{document}